\documentclass[%
 aip,
 amsmath,amssymb,
 aps,]{revtex4-1}

\usepackage{amsmath}
\usepackage{amssymb}
\usepackage{graphicx}
\usepackage{epsfig}
\usepackage{color}
\usepackage{fullpage}
\usepackage{subfig}
\usepackage{bbm}

\newcommand{\du}{\, \mathrm{d}}

\begin{document}

\title{Bouncing droplets on a billiard table}

\author{David Shirokoff}
 \email{david.shirokoff@mail.mcgill.ca}
\affiliation{Department of Mathematics and Statistics\\
 McGill University, Montreal, QC  H3A0B9, CAN}
\date{\today}

\begin{abstract}
In a set of experiments, Couder et. al. demonstrate that an oscillating fluid bed may propagate a bouncing droplet through the guidance of the surface waves.  We present a dynamical systems model, in the form of an iterative map, for a droplet on an oscillating bath.  We examine the droplet bifurcation from bouncing to walking, and prescribe general requirements for the surface wave to support stable walking states.  We show that in addition to walking, there is a region of large forcing that may support the chaotic motion of the droplet.   Using the map, we then investigate the droplet trajectories in a square (billiard ball) domain.  We show that in large domains, the long time trajectories are either non-periodic dense curves, or approach a quasiperiodic orbit.  In contrast, in small domains, at low forcing, trajectories tend to approach an array of circular attracting sets.  As the forcing increases, the attracting sets break down and the droplet travels throughout space.
\end{abstract}

\maketitle

\textbf{In a recent experiment \cite{CouderFortGautierBoudaoud2005}, Couder et. al. demonstrate that under sufficient conditions, a fluid droplet will bounce infinitely often on an oscillating bed of the same fluid.  Moreover, upon increasing the acceleration of the fluid bed, the droplet will transition \cite{CouderFort2006, CouderProtiereFortBoudaoud2005, ProtiereBoudaoudCouder2006} from a stable bouncing state to a self propagating one.  The resulting coupling between the droplet motion and underlying wave field has led to a variety of experiments \cite{CouderFort2006, CouderProtiereFortBoudaoud2005, EddiBoudaoudCouder2011, EddiDecelleFortCouder2009, EddiFortMoisyCouder2009, EddiSultanMoukhtarFortRossiCouder2011, EddiTerwagneFortCouder2008, FortEddiBoudaoudMoukhtarCouder2010, GiletBush2009, OzaRosalesBush2011, ProtiereBoudaoudCouder2006, HarrisMoukhtarFortCouderBush} demonstrating that the droplet trajectories may exhibit wave-like characteristics.  Although in some of the experiments, the droplet trajectories appear to approach stable structures, such as quasiperiodic orbits, in other situations, the trajectories appear ergodic or must be understood in a statistical sense.  The goal of this paper is to describe the behavior of droplet trajectories, with an eye towards understanding their statistical behavior.  We do this by adopting a dynamical systems approach, and model the motion of the droplets using a discrete iterative map.  In particular we show that a map retaining one past position of the droplets history is simple enough to yield analytic results, yet complex enough to reproduce a variety of droplet behaviors: including the bifurcation from bouncing to walking, as well as quasiperiodic orbits and the appearance of dense trajectories in bounded domains.}

\section{Introduction}

In the experimental setup described in the previous paragraph, bouncing droplets on a fluid bed may couple with the underlying wave field to create a combined particle-wave system.  Upon each bounce with the fluid bed, the droplet generates a wave field in the bath.  As the droplet continues to bounce on the bed, the previously generated waves act to guide the trajectory of the particle.  Following \cite{CouderFort2006, ProtiereBoudaoudCouder2006} we refer to the combination of the moving droplet dressed with a local wave as a \emph{walker}.  In many cases, the underlying wave field may also act as a medium to guide droplets in the vicinity of domain boundaries or even other droplets
(see also \cite{Bush2010} for a review).  For instance, in the first of a series of experiments \cite{CouderProtiereFortBoudaoud2005, ProtiereBoudaoudCouder2006}, the authors show that two droplets may orbit or scatter without direct contact, but rather through the mediation of the underlying fluid bath.  In addition, \cite{ProtiereBoudaoudCouder2006} goes further to investigate the phase diagram for bouncing and walking droplets with different sizes and accelerations of the fluid bed, as well as providing the first steps towards a phenomenological model for the droplet wave system.  In their model, the authors average over one droplet interaction to obtain an ODE describing droplet motion.  A more detailed understanding of the droplet interactions and motion can also be found in \cite{MolacekBush1, MolacekBush2}.

Several recent experiments have also examined the trajectories of bouncing droplets in bounded domains.  For instance, when a droplet is sent through a single slit scattering
experiment \cite{CouderFort2006}, the droplet may propagate with an apparently random scattering angle.  Upon repeating the experiment for many droplets, the data shows that the droplets propagate with a probability distribution statistically analogous to a scattered wave amplitude.  Meanwhile, in \cite{EddiFortMoisyCouder2009} Eddi et al. examine the motion of droplets in a confined billiard setting.  In analogy with quantum tunneling, they show that under sufficient conditions, instead of reflecting off the boundary walls, the droplets may occasionally cross a \emph{dead zone} region which does not support stable walking trajectories.  The paper also experimentally shows that one may obtain stable quasiperiodic trajectories, or the appearance of ergodic trajectories for droplets in a billiard domain.  Further experiments\cite{HarrisMoukhtarFortCouderBush} examine the statistical nature of the droplet position within a confined domain. Lastly, \cite{FortEddiBoudaoudMoukhtarCouder2010} examine the trajectory of bouncing droplets in a rotating fluid.  In a qualitative analogy with Landau orbits, they show that at sufficient forcing of the fluid bed, the droplets move in circular orbits with quantized radii.  The papers \cite{CouderFort2006, FortEddiBoudaoudMoukhtarCouder2010} also provide a simplified phenomenological model for the droplet motion and numerically reproduce the qualitative behavior of the droplet trajectories.  One drawback of the model is that the fluid wave behavior is imposed as an ansatz without a model for the evolution of the fluid bath.  Subsequent theoretical improvements were included in \cite{EddiSultanMoukhtarFortRossiCouder2011} where the authors provide a more detailed description of the waves generated near the Faraday threshold.

The experiments have not only been limited to the behavior of single droplets, but also include many droplet systems.  For instance, \cite{EddiTerwagneFortCouder2008} demonstrate that droplet pairs may form localized, orbiting bound states, and even small crystalline structures.  Further work \cite{EddiDecelleFortCouder2009} shows that the crystalline structures may include many different Archimedean lattices, while in \cite{EddiBoudaoudCouder2011}, Eddi et al. examine the instabilities in periodic hexagonal and square arrays.

In the first section of our paper, we introduce an iterative map model for the droplet trajectories.  In section (II) we outline a gravity-capillary model, similar to the one introduced in \cite{EddiSultanMoukhtarFortRossiCouder2011}, for the underlying fluid bath.  Using the wave model, we show that the droplet motion is primarily a result of the most recent bounce, and then analytically examine the bouncing to walking bifurcation.  We show that the sum of past droplet impacts create an outgoing standing wave, and that the model also predicts a mechanism for the transition to chaotic droplet motion.  Lastly, we examine droplet trajectories in bounded, square domains.  In the first case, we examine domains much larger than the fluid wavelength and show the existence of either dense trajectories or quasiperiodic orbits.  In the second case, we examine small domains and show that trajectories may bounce between different regions of space.

\section{The Iterative map}

In this paper we are interested in understanding the two dimensional dynamics of the fluid droplets, as they propagate around the oscillating bath.  As a result, we work in two spatial dimensions and record only the two dimensional position of the droplet, thereby ignoring the vertical (bouncing) motion of the particle.
Let $\Omega \subset \mathbbm{R}^2$ be the two dimensional domain of the fluid bath, and denote the continuous time position and velocity of the fluid droplet by $\mathbf{y}(t) \in \Omega$ and $\mathbf{v}(t)$ respectively.  In addition, we let $h(\mathbf{x}, t; \mathbf{y})$ denote the surface height of the fluid bath at $\mathbf{x}$.  Since the wave field depends on the past history of the particle, we include $\mathbf{y}(\tau)$ (for $\tau < t$) as a functional parameter in the wave field.

To derive a set of dynamic equations,  we adopt a simplified phenomenological model for the contact interaction of the fluid bed and droplet.  In doing so we do not resolve the microscopic interaction between the fluid bed and droplet, but rather assume a spontaneous interaction.  Specifically, we make the following simplifying assumptions pertaining to the interaction:
\begin{enumerate}\label{Assump_1}
	\item [A1.] The nondissipative forcing from the fluid bath on the particle is proportional to the wave field slope at the time and location of contact. \label{Assump_2}
	\item [A2.] At each bounce, the particle provides a point source forcing to the fluid wave field.
\end{enumerate}
We note that the general assumptions (A1)--(A2) are also made in the phenomenological model taken in \cite{CouderFort2006, FortEddiBoudaoudMoukhtarCouder2010}.  In general, the droplet size (droplets $\leq 1mm$ in diameter) can result in large variations in the dynamics, however (A1)--(A2) simplifies them to be point particles.
Following assumption (A1), the equations of motion for the fluid droplet take the form
\begin{eqnarray} \label{Motion1}
	\dot{\mathbf{y}} &=& \mathbf{v} \\  \label{Motion2}
	\dot{\mathbf{v}} &=& -\big[F \nabla h(\mathbf{x}, t; \mathbf{y}) \big|_{\mathbf{x} = \mathbf{y}} + \gamma \mathbf{v}^- \big] \delta_p(t),
\end{eqnarray}
where $F$ is the amplitude of the forcing on the particle, and $\delta_p(t) = \delta_p(t + T)$ is the periodic Dirac delta function.  To capture dissipation in the droplet-bath interaction, we have added the additional term, $\gamma \mathbf{v}^- \delta_p(t)$, where $\mathbf{v}^-$ is understood to be the velocity prior to impact.  Mathematically, we take $\mathbf{v}^-(t) \delta_p(t) = \mathbf{v}(t - \epsilon) \delta_p(t)$ with $\epsilon \rightarrow 0$, to correctly define the product of a distribution $\delta(t)$ with a discontinuous function $\mathbf{v}$.  One should note that in the general case, the shape of the wave field $h(\mathbf{x}, t; \mathbf{y})$, depends on the previous history of the particles position $\mathbf{y}(t)$.

Since the forcing is periodic, we may integrate the equations of motion (\ref{Motion1})--(\ref{Motion2}) over one period to obtain a discrete map. To compactify the notation, we first choose the period of bouncing (T) as the natural time scale, and let $\mathbf{y}_n = \mathbf{y}(n + \epsilon)$ and $\mathbf{v}_n = \mathbf{v}(n + \epsilon)$ with ($\epsilon \rightarrow 0$), denote the droplets position immediately following the $nth$ bounce.  Upon integrating equations (\ref{Motion1})--(\ref{Motion2}), we obtain the iterative map
\begin{eqnarray} \label{Map_y}
	\mathbf{y}_{n+1} &=& \mathbf{y}_n + \mathbf{v}_n  \\ \label{Map_v}
	\mathbf{v}_{n+1} &=& (1-\gamma) \mathbf{v}_n - F \nabla h(\mathbf{x}, n+1; \mathbf{y}_n, \mathbf{y}_{n-1}, \ldots)\big|_{\mathbf{x} = \mathbf{y}_{n+1}}.
\end{eqnarray}
In the special case when $\gamma = 0$, and $h = h(\mathbf{x})$ is independent of the droplets history, then (\ref{Map_y})--(\ref{Map_v}), reduce to the \emph{standard map} \cite{Chirikov1979}.  The standard map is the discrete analogue of a particle moving in a strobed Hamiltonian system, and consequently the map preserves phase-space volume.  We note that the presence of a path memory, with or without dissipation, breaks the discrete analogue of Liouville's theorem, thereby allowing stable attractors in phase space.

Thus far, we have not explicitly stated a model for the evolution of the fluid bath.  Hence at this point one could take a variety of models for $h(\mathbf{x}, t; \mathbf{y})$ that accurately capture various properties of the underlying fluid field.

\section{A model for the wave field}

In the following section, we introduce gravity-capillary waves \cite{LandauLifshitzFluids} as a model for the fluid wave evolution.  We show that the most recent droplet impact dominates the contribution to walking, while including many previous impacts creates an outgoing standing wave centered about the droplet.  We then use the approximation to analytically examine the resulting bifurcation.  Lastly, we examine the effects of dissipation on the walking velocity.

In the linear theory of gravity-capillary waves, one assumes an irrotational velocity field and models the fluid with a velocity potential $\phi(\mathbf{x}, z, t)$ with $\mathbf{x} \in \Omega$.  Here, $z$ is the vertical direction and $z = 0$ is aligned with the unperturbed surface height $h = 0$.  It then follows that $h(\mathbf{x}, t)$ and $\phi(\mathbf{x}, z, t)$ satisfy the linearized gravity-capillary equations.  Namely, $\phi(\mathbf{x}, z, t)$ is a harmonic function which vanishes at $z \rightarrow -\infty$
\begin{equation}
   \begin{array}{rcll}
		\Delta \phi & = & 0 \quad & \textrm{for}\;\; z < 0
        \\ \rule{0ex}{2.5ex}
		\phi & = & 0
        \quad & \textrm{for}\;\; z \rightarrow -\infty.
        \rule{0ex}{2.5ex}
   \end{array}
\end{equation}
In addition, $\phi(\mathbf{x}, z, t)$ is coupled to $h(\mathbf{x}, t)$ via the kinetic and dynamic boundary conditions at $z = 0$
\begin{eqnarray} \label{Kinetic_BC}
	h_t & = & \phi_z \\ \label{Dynamic_BC}
	\phi_t + g h - \frac{\sigma}{\rho} \Delta_2 h + 2 \frac{\nu}{\rho} \phi_{zz} & = & -\sum_{j = 0}^{n-1} \frac{f_0}{\rho} \delta(\mathbf{x} - \mathbf{y}_{n-j}) \delta(t - t_{n-j}).
\end{eqnarray}
Here $\Delta_2$ is the 2D Laplacian in $\mathbf{x}$, while $\Delta$ is the 3D Laplacian in $(\mathbf{x}, z)$.  Meanwhile, the point source forcing at time $t_n = n$ enters as a delta function, via assumption (A2), to model the instantaneous interaction with the bed.  We remark that the dissipative term in equation (\ref{Dynamic_BC}) is only an effective dissipation as the assumption of an irrotational fluid field breaks down in a small viscous boundary layer near the wave field surface \cite{EddiSultanMoukhtarFortRossiCouder2011, KumarTuckerman1994, LandauLifshitzFluids}.

To nondimensionalize the equations, we again use the period of bouncing ($T$) as the time scale, and take a length scale \cite{LengthScale} set by the pure gravity waves as $L = g T^2$.  Letting, $\mathbf{x} \rightarrow L \mathbf{x}$, $t \rightarrow T t$, we also let $\phi \rightarrow (\frac{L^2}{T}\phi )(\frac{f_0 T}{\rho L^4})$, $h \rightarrow (L h)(\frac{f_0 T}{\rho L^4})$ and $F \rightarrow (\frac{L}{T}F)(\frac{\rho L^4}{f_0 T})$, where the dimensionless factor $(\frac{f_0 T}{\rho L^4})$ is included to simplify the equations to
\begin{eqnarray} \label{DimensionlessEq1}
	h_t &=& \phi_z \\ \label{DimensionlessEq2}
   \phi_t +  h - B^{-1} \Delta_2 h + 2\mu \phi_{zz}  &=& - \sum_{j = 0}^{n-1} \delta(\mathbf{x} - \mathbf{y}_{n-j})\delta(t - t_{n-j}).
\end{eqnarray}
Here $\mu = \nu/ (\rho g^2 T^3)$ is the dimensionless viscosity, while $B = g\rho L^2/\sigma = g^3 \rho T^4/\sigma$ is the ratio of buoyancy to surface tension restoring forces in the wave field.   Typical experimental values \cite{ProtiereBoudaoudCouder2006}, are $\sigma = .0209 Nm^{-1}$, $\rho = .965 \times 10^3 kg$ $m^{-3}$, $T = 25^{-1}s$, $\nu = 5 \times 10^{-3} Pa s$ to $0.1 Pa s$ yielding $B \sim 120$, $L \sim 16mm$ and $\mu \sim .001$ to $.016$.  For calculations we will typically take $\mu = 0.008$, which is the midpoint of the viscosity range.

Along with rest conditions at $t_0$
\begin{eqnarray} \label{InitialData1}
	\phi(\mathbf{x}, z, t_0) &=& 0 \\ \label{InitialData2}
	h(\mathbf{x}, t_0) &=& 0
\end{eqnarray}
we take equations (\ref{DimensionlessEq1})--(\ref{DimensionlessEq2}) to describe the fluid evolution between bounce $n$ and $n+1$.

The map depends on four parameters, the dissipation of the wave ($\mu$), the dissipation of the droplet bounce ($\gamma$), the shape of the dispersion relation ($B$) and the acceleration or forcing of the particle ($F$).  In the subsequent sections, we study (\ref{DimensionlessEq1})--(\ref{DimensionlessEq2}) to understand the resulting particle motion for various forcing and domains.

\subsection{Solution in free space} \label{FreeSpace}

In this section we construct the iterative map (\ref{Map_y})-(\ref{Map_v}) with (\ref{DimensionlessEq1})--(\ref{DimensionlessEq2}) in free space ($\Omega = \mathbbm{R}^2$).  We do this by solving for the wave field $h(\mathbf{x}, t; \mathbf{y}_n)$ from one point source interaction at time $t_n$:
\begin{eqnarray} \label{IRDimensionlessEq1}
	h_t &=& \phi_z \\ \label{IRDimensionlessEq2}
   \phi_t +  h - B^{-1} \Delta_2 h + 2\mu \phi_{zz}  &=& -\delta(\mathbf{x} - \mathbf{y}_{n})\delta(t - t_{n}).
\end{eqnarray}

Linear superposition then allows one to add the contributions from many past bounces.  With the exception of adding dissipation, the solution follows very closely that of the standard Cauchy-Poisson problem for forced gravity-capillary waves.   Specifically, we seek an expansion for $\phi$ of the form ($k = |\mathbf{k}|$)
\begin{eqnarray} \label{Ansatz}
    \phi(\mathbf{x}, z, t) = \int A(\mathbf{k}, t) e^{k z} e^{\imath \mathbf{k}\mathbf{x}} \du \mathbf{k}.
\end{eqnarray}
Upon substituting the ansatz (\ref{Ansatz}) into equations (\ref{DimensionlessEq1})--(\ref{DimensionlessEq2}), one obtains an initial value problem for each $A(\mathbf{k}, t)$.  The solution for $h(\mathbf{x}, t; \mathbf{y}_n)$ over $t > t_n$ then becomes
\begin{eqnarray} \label{GCPointSource}
    h(\mathbf{x}, t; \mathbf{y}_n) &=& - \int \frac{k}{2\pi \omega_D} \sin(\omega_D (t-t_n) ) e^{\imath  \mathbf{k}(\mathbf{x} - \mathbf{y}_n)} e^{-\mu k^2 (t-t_n)} \du \mathbf{k},
\end{eqnarray}
where $\omega_D^2 = (k + B^{-1} k^3) - \mu^2 k^4$ is the dispersion relation.  Note that in the case when $\omega_D^2 < 0$, the function $\frac{\sin(\omega_D)}{\omega_D} e^{-\mu k^2}$ becomes a strict exponential decay, corresponding to overdamping  of the large modes.

To compute the field, we first integrate over the angular component $\theta$ in (\ref{GCPointSource}) to obtain the impulse response
\begin{eqnarray} \label{Simplified_h}
	h(\mathbf{x}, t; \mathbf{y}_n) &=& h_{0}(|\mathbf{x} - \mathbf{y}_n|, t - t_n)	\\
h_{0}(r, \tau) &=& -\int_{0}^{\infty} \frac{k^2}{\omega_D} \sin(\omega_D \tau) J_0(k r) e^{-\mu k^2 \tau} \du k 	
\end{eqnarray}
Here $J_0(x)$ is the zeroth order Bessel function.  As a one-dimensional integral, we may numerically evaluate (\ref{Simplified_h}) and compute the iterative map
\begin{eqnarray} \label{Update_y}
	\mathbf{y}_{n+1} &=& \mathbf{y}_n + \mathbf{v}_n  \\ \label{Update_v}
	\mathbf{v}_{n+1} &=& (1-\gamma) \mathbf{v}_n + F \mathbf{g}(\mathbf{y}_{n+1}, \ldots \mathbf{y}_1) \\ \label{Update_Map}
    \mathbf{g}(\mathbf{y}_{n+1}, \dots, \mathbf{y}_1) &=& - \sum_{j = 0}^{n-1} \nabla h_0(r_{n-j}, j+1)\big|_{\mathbf{x} = \mathbf{y}_{n+1}}.
\end{eqnarray}
where for brevity we have let $r_n = |\mathbf{x} - \mathbf{y}_n|$.

\begin{figure}
  \centering
  \subfloat[Gravity-capillary wave field at times $T = 1, 2,3$.]{\label{WaveFieldFreeSpace}\includegraphics[width=0.5\textwidth]{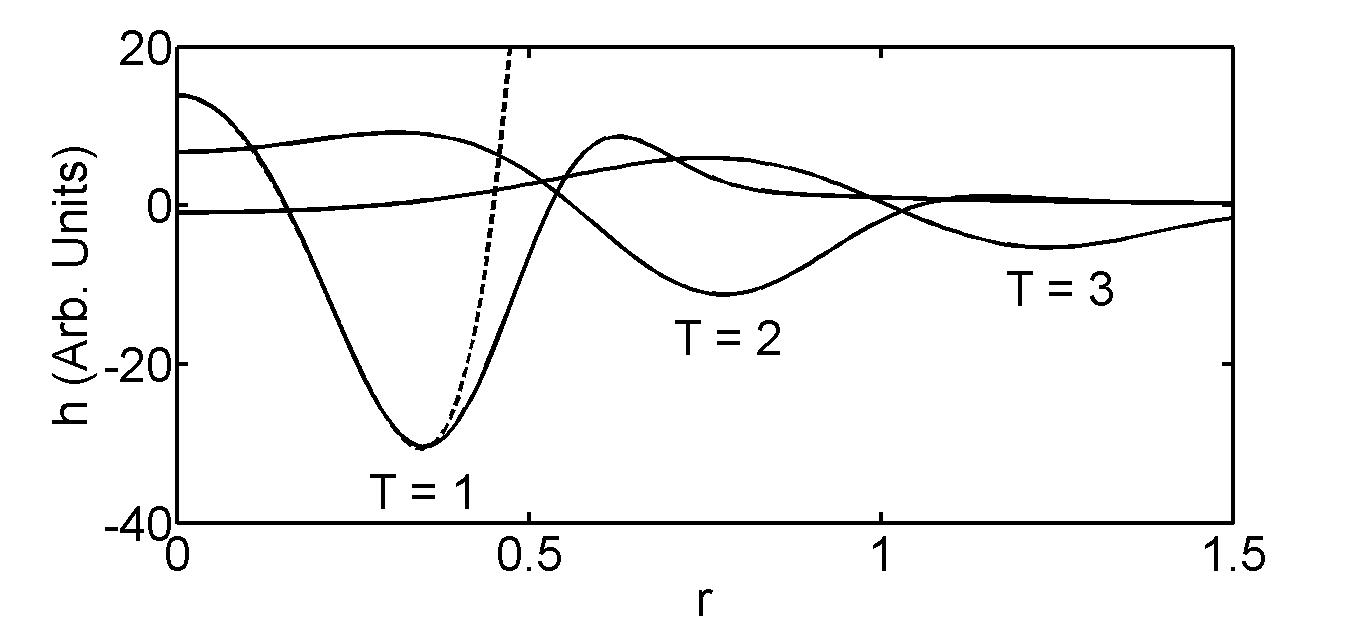}}
  \subfloat[Poincar\'{e} map of (\ref{VelocityMap_FreeSpace})-(\ref{Polynomial}) and $F > F_{crit}$.]{ \label{PoincareMapFreeSpace}\includegraphics[width=0.5\textwidth]{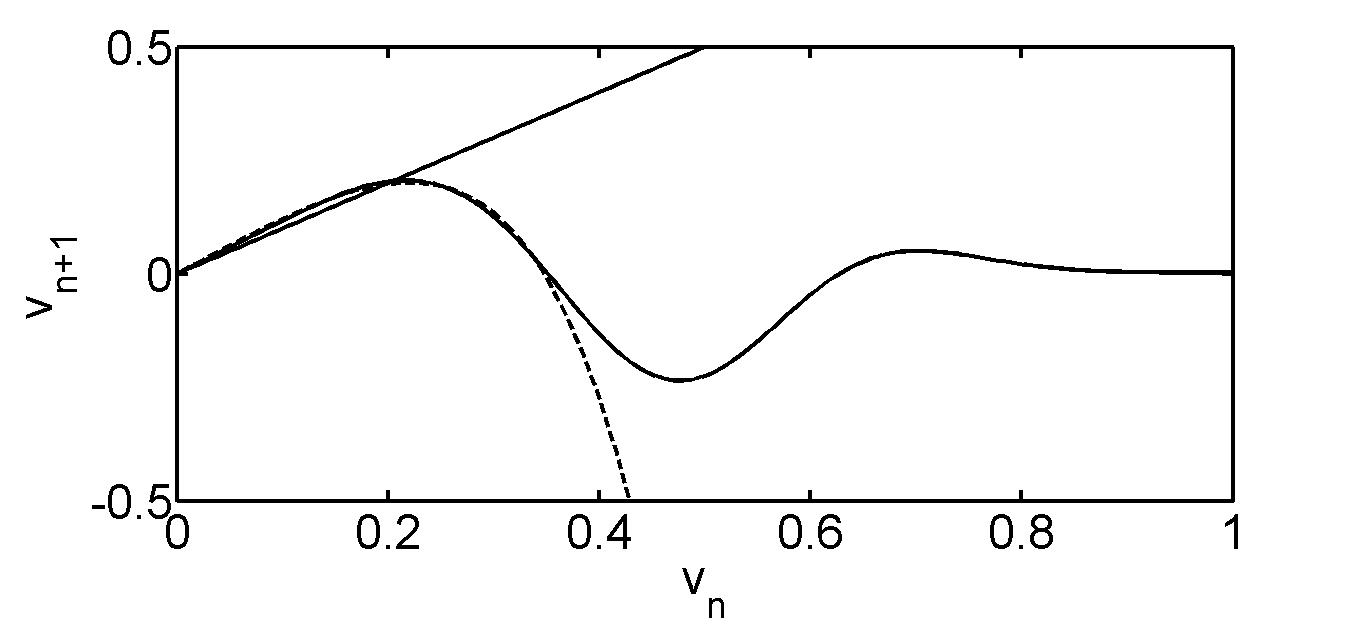}}
  \caption{Shows the radial wave field impulse response $h_0(r, T)$ at different times, and the Poincar\'{e} map for gravity-capillary model ($B = 120$, $\mu = 0.008$). The dashed line shows the polynomial fit (\ref{Polynomial}). }
\end{figure}

To extract a simplified expression for the map, we now focus on computing $h_0(r, \tau)$ for the physically relevant parameters $B = 120$ and $\mu = 0.008$.  Here figure (\ref{WaveFieldFreeSpace}) shows the radial wave field $h_0(r, \tau)$ for different $\tau$.  In particular the point source wave both disperses and radiates outward.  Within $\tau \geq 3$ periods, the wave has traveled several wavelengths and has minimal support near $r = 0$. At $\tau = 2$ and values of $r < 0.4$, the contribution of the wave $\partial_r h_0(r, 2)$ to the iterative map $(\ref{Update_Map})$ is small compared to $\partial_r h_0(r, 1)$.  We therefore neglect the second impact and make the following assumption
\begin{enumerate}\label{Assump_3}
	\item [A3.] The droplet walking dynamics depend only on the most recent droplet bounce. Explicitly, the assumption yields
	\begin{eqnarray} \label{Map_A3}
		\mathbf{g}(\mathbf{y}_{n+1}, \mathbf{y}_n) = - \nabla h_0(r_n, 1)\big|_{\mathbf{x} = \mathbf{y}_{n+1}}.
	\end{eqnarray}
\end{enumerate}

The assumption (A3) is most valid provided the forcing $F$ is far below the Faraday threshold.  Near the Faraday threshold, the equations (\ref{Kinetic_BC})--(\ref{Dynamic_BC}) no longer accurately describe the wave field. Instead, one must incorporate the periodic forcing of the bed into a model for $h(\mathbf{x}, t)$.  Such an inclusion results in two effects: i) a different shape of the wave field radiating from point sources, namely one with a fixed wavelength, ii) a strong memory of past bounce locations.

	Although the Faraday threshold does not enter into the current model, using known experimental data, we may estimate the valid region of forcing $F$.  Experiments \cite{EddiSultanMoukhtarFortRossiCouder2011} show that Faraday wave memory effects become important when the bed acceleration $\tilde{\gamma}_b$ is close to $\tilde{\gamma}_f$.  Specifically, data collected for a bouncing period of $T = 40$ suggests the crossover occurs when $(\tilde{\gamma}_f - \tilde{\gamma}_b )/\tilde{\gamma}_f$ is somewhere between $(0.07, 0.17)$.  For this approximation, we take the crossover to be at $10^{-1}$.  Experiments \cite{EddiSultanMoukhtarFortRossiCouder2011, ProtiereBoudaoudCouder2006} have also measured the walking and Faraday threshold accelerations of the bed at $\tilde{\gamma}_w \sim 3.75 g$ and $\tilde{\gamma}_f \sim 4.5g$ for $T = 25^{-1}$ and $\tilde{\gamma}_w \sim 3.2 g$ and $\tilde{\gamma}_f \sim 4.1g$ for $T = 40^{-1}$.  The walking accelerations may vary depending on droplet size, however typical values for fixed $T = 25^{-1}$ are between $3.1g$ and $3.8g$.  Finally, we note that $F \propto \tilde{\gamma}_b$, so that $F/F_{crit} = \tilde{\gamma}/\tilde{\gamma}_w$. Estimating the maximum value of $F/F_{crit}$ then yields
	\begin{eqnarray}
		1 - \frac{\tilde{\gamma}}{\tilde{\gamma}_f} &\geq& 10^{-1}\\
		\frac{\tilde{\gamma}_w}{\tilde{\gamma}_f}\frac{F}{F_{crit}} &\leq& \frac{9}{10}.
	\end{eqnarray}
	We therefore expect that parametric effects are important when $F/F_{crit}$ is larger than $1.1$ - $1.2$.

With the simplifying assumption (A3), we may reduce the map (\ref{Update_y})-(\ref{Update_Map}) to one-dimension, and analytically examine the bifurcation from bouncing to walking.  To do so, align the droplet position and velocity with the x-axis $\mathbf{y}_n = r_n \hat{\mathbf{x}}$, $\mathbf{v}_n = v_n \hat{\mathbf{x}}$ to obtain
\begin{eqnarray}  \label{VelocityMap_FreeSpace}
    v_{n+1} &=& (1-\gamma)v_n + F g(v_n) \\ \label{MapIntegral}
    g(v) &=& \int_{0}^{\infty} \frac{k^3}{\omega_D} \sin(\omega_D) J_0'(kv)  e^{-\mu k^2 } \du k\\
\label{Polynomial}
	g(v) &\approx&  1150 v (1 - 4.32 v^2 + 38.86 v^4)
\end{eqnarray}
The last line (\ref{Polynomial}) is an approximate polynomial fit for $g(v)$, while $J_0'(z) = \frac{d}{dz} J_0(z)$.   We note that (even regardless of $B$ and $\mu$) $v_n = 0$ is a fixed point of the map (\ref{VelocityMap_FreeSpace}), and therefore bouncing droplets with a fixed location $r_{n+1} = r_n$ are always solutions of (\ref{VelocityMap_FreeSpace}).  To illustrate the nature of the bifurcation from stable bouncing to walking, we fix a value of $\gamma$ and continually vary $F$ as the bifurcation parameter.
Here figures (\ref{WaveFieldFreeSpace}) and (\ref{PoincareMapFreeSpace}) show the wave field and Poincar\'{e} map, with the associated polynomial fit (\ref{Polynomial}).

\begin{figure}[htb!]
  \centering
  \subfloat[Bifurcation diagram.]{\label{Polynomial_Bifurcation}\includegraphics[width=0.5\textwidth]{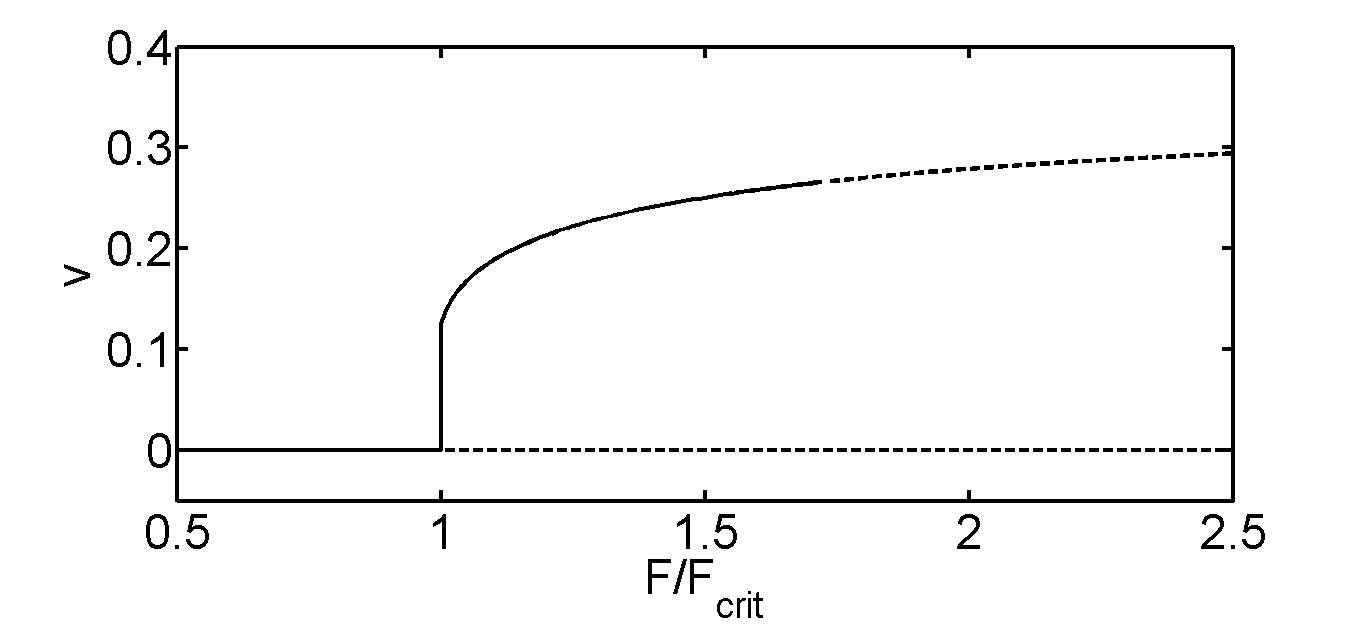}}
  \subfloat[Droplet velocity versus forcing.]{ \label{Polynomial_VelocityOrbits}\includegraphics[width=0.5\textwidth]{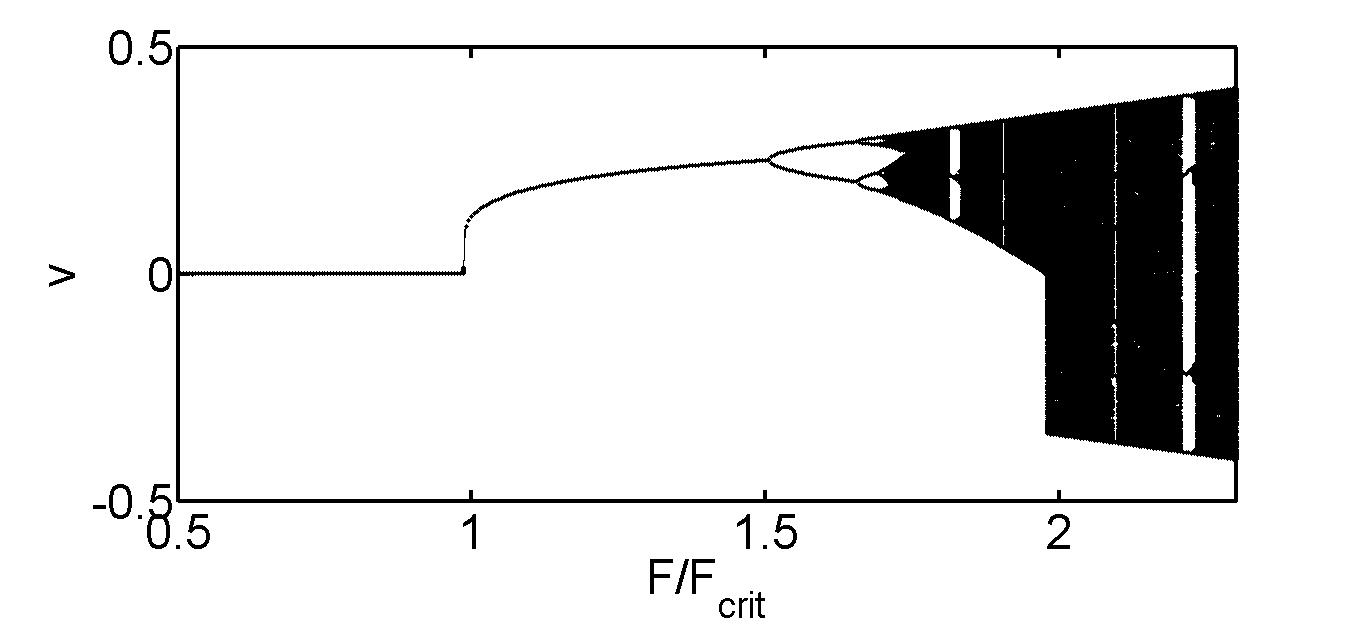}}
  \caption{Shows the bifurcation from stable bouncing to walking for the gravity-capillary model (\ref{VelocityMap_FreeSpace})-(\ref{Polynomial}). In (a), the transition is a supercritical pitchfork where $F_{crit} = 1150^{-1}$, $B = 120$, $\mu = 0.008$. In (b), after the initial bifurcation, the droplet undergoes a transition to chaos.}
\end{figure}

As one increases $F$, the fixed point solution $v_n = 0$ becomes unstable at which point the system undergoes a pitchfork bifurcation.  For values of $F$ above the critical forcing ($F > F_{crit} = 1150^{-1}$), the points $v_n$ converge to a new fixed point solution $v^*$, indicating a bifurcation from stable bouncing to walking.  For instance figure (\ref{Polynomial_Bifurcation}) shows the bifurcation diagram in the case when $\gamma = 1$.  As one further increases $F$, the stable walking solution bifurcates a second time into a two-period orbit, followed by a transition to chaos (figure (\ref{Polynomial_VelocityOrbits})).  We remark that the pitchfork bifurcation and transition to chaos occur for a large range of $\mu$ and $B$, however the supercritical bifurcation seen in (\ref{Polynomial_Bifurcation}) is not generic.  For instance, the supercritical bifurcation to walking is due to the sign of the derivative $h^{(4)}(0) = -g^{(3)}(0) > 0$.  In general, by varying $B$ (ie. the shape of the wave) one may realize both positive and negative values of $h^{(4)}(0)$, where $h^{(4)}(0) > 0$ implies a supercritical bifurcation while $h^{(4)}(0) < 0$ implies a subcritical one.  Experiments have observed both sub and supercritical bifurcations, however the difference may arise from other effects not considered in the current model, such as instabilities in the vertical droplet dynamics.

The iterative map also provides predictions for the walker velocity and wave velocity which we now compare to experimental data.  First, in the gravity-capillary wave model (\ref{IRDimensionlessEq1})-(\ref{IRDimensionlessEq2}), the point source forcing excites all wave lengths of $h(\mathbf{x}, t)$.  Hence, the wave has a minimum group velocity $v_{g}^{min} = 0.33 = 132mm$ $s^{-1}$, obtained by the gravity-capillary dispersion relation, which approximately limits the speed of the disturbance.  Meanwhile, the fixed point walking velocities are $v^* \sim 0.25 = 100mm$ $s^{-1}$, while the characteristic standing wavelength is $\lambda = 0.6 = 9.6mm$.  Experiments (figure 6a. in \cite{EddiSultanMoukhtarFortRossiCouder2011}) show a wave propagating roughly $20mm$ in a time of $\sim 0.2 s$ for a minimum group velocity of $\tilde{v}_g^{min} \sim 100mm$ $s^{-1}$, while the standing wavelength $\tilde{\lambda} = 4.75mm$ .  Although the gravity-capillary wave is in good agreement with the fluid experiments, the maximum \cite{ProtiereBoudaoudCouder2006} experimental droplet velocity is roughly a factor of 5 smaller: $\tilde{v}^* = 20mm$ $s^{-1}$.  Despite yielding a qualitative agreement, the iterative map model over estimates the droplet velocity by locking the droplet velocity to the wave.  Here the discrepancy is a result of the simplified assumptions (A1)-(A2).  In particular, a detailed model for the droplet bouncing dynamics and surface interactions may account for the velocity mismatch.

\begin{figure}[htb!]
  \centering
   \subfloat[Droplet velocity versus forcing.]{ \label{Polynomial_VelocityOrbitsFull}\includegraphics[width=0.5\textwidth]{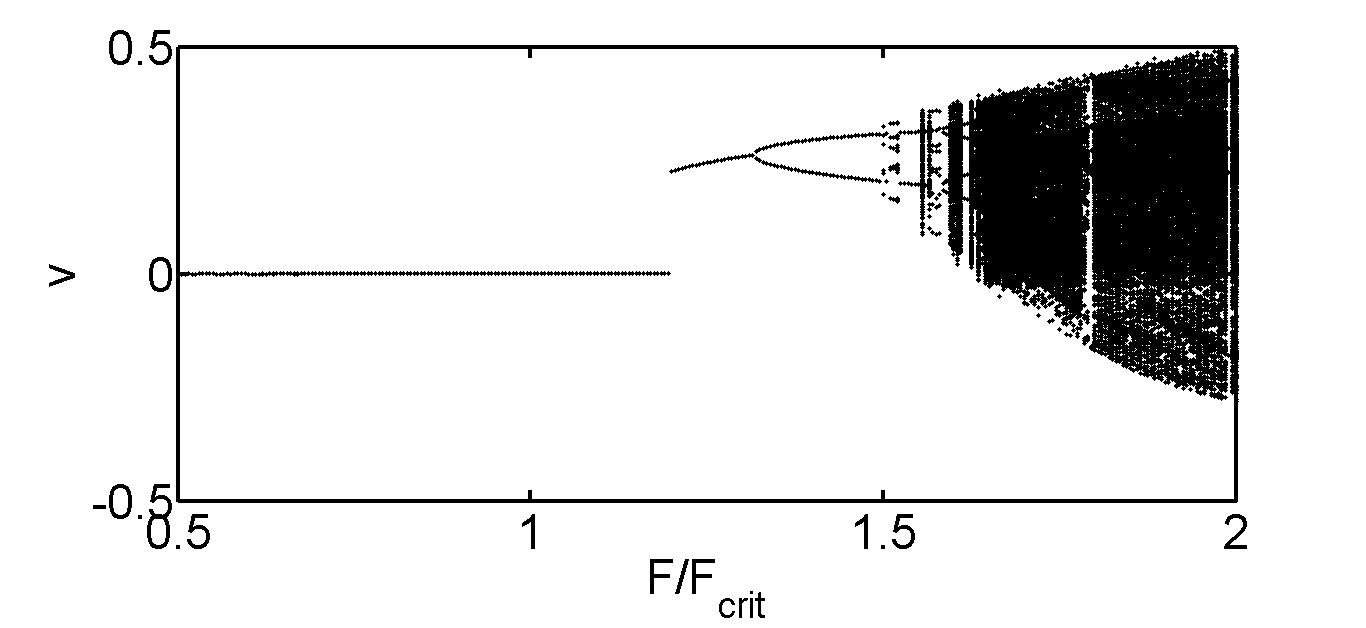}}
	\subfloat[Fully developed wave.]{\label{FullWaveField}\includegraphics[width=0.5\textwidth]{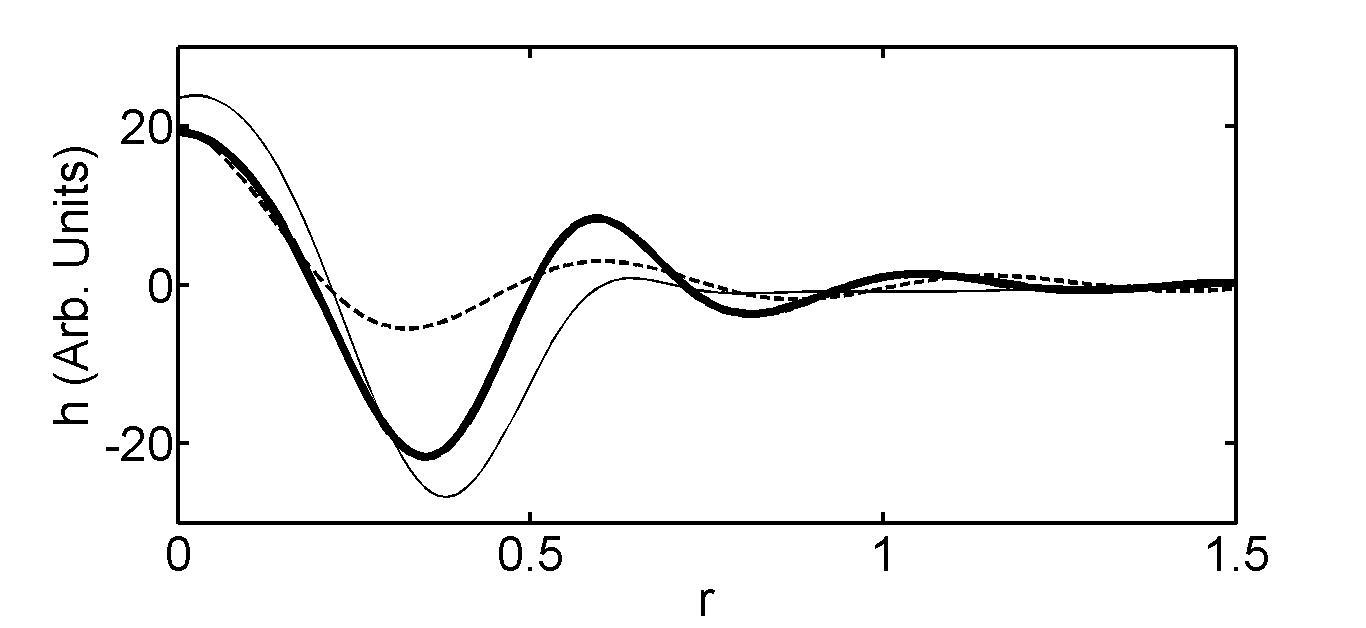}}
\caption{  (\ref{Polynomial_VelocityOrbitsFull}) Shows the droplet velocity including 10 previous bounces, where $F_{crit} = 1150^{-1}$ is the single bounce critical forcing.  Multiple bounces can increase the threshold for walking and also lower the critical forcing for chaos. (\ref{FullWaveField}) Shows the fully developed wave field for a walking droplet as a superposition of shifted sources $\sum_{n = 1}^{10} h_0(r_n, n)$ where $r_n = r + (n-1) v_n$ and $v_n = 0.05$ (thick line), $v_n = 0.25$ (thin line). The dashed line is an approximate Bessel function wave field used in section (V). The primary contribution to walking comes from the most recent bounce.}
\end{figure}

Lastly, we examine the effects of including multiple bounces in the wave field. Here figure (\ref{Polynomial_VelocityOrbitsFull}) shows the velocity dynamics of including 10 past bounces, while (\ref{FullWaveField}) shows the fully developed wave field for a walking droplet.  Upon the onset of walking, the superposition of many past impacts creates an apparent standing wave pattern (\ref{FullWaveField}), which qualitatively agrees with experiments\cite{EddiSultanMoukhtarFortRossiCouder2011, ProtiereBoudaoudCouder2006}.  Explicitly, the wave field may be written as $\sum_{n = 1}^{10} h_0(r_n, n)$ where $r_n = r + (n-1) v^*$ and $v^*$ is the fixed point velocity.

Figure (\ref{Polynomial_VelocityOrbitsFull}) also shows that including multiple bounces can increase the threshold for walking and lower the relative forcing required for chaos.  Including multiple bounces, however, only mildly changes the nature of the transition to chaos.  Experimentally observing such a transition requires measuring variations in the droplet velocity.  For instance, the droplet transitions from a steady velocity $v^*$, to one that jumps with alternating step size $v^* \pm \epsilon$. Experimentally, one would observe a time averaged mean velocity $v^*$ (indicating no change), while precise measurements would detect small periodic variations.  We note that the transition to chaos described here does not account for possible instabilities in the angular droplet dynamics.

\subsection{Varying dissipation}

Although we have been using the forcing $F$ as a bifurcation parameter, the viscosity of the fluid $\mu$ can also vary depending on the vibration of the bed.  In this section we examine the effect of varying dissipation in model (\ref{VelocityMap_FreeSpace})--(\ref{MapIntegral}).  We first remark that one may asymptotically approximate $h(\mathbf{x}, t_{n+1}; \mathbf{y}_n)$ for $\mu \gg 1$.  The result is an over-damped wave, which does not support the steady walking of droplets.  For large $\mu$, the terms inside the integral may be approximated as follows: when $k > O(\mu^{-2/3})$, the value $\mu^2 k^4 > \omega_0^2$ at which point
\begin{eqnarray}
	\frac{\sin(\omega_D)}{\omega_D} e^{-\mu^2 k^4} &=& \frac{\sinh( (\mu^2 k^4 -k - B^{-1} k^3)^{1/2} )}{(\mu^2 k^4 -k - B^{-1} k^3)^{1/2} } e^{-\mu^2 k^4} \\
	&\approx& \frac{1}{\mu k} e^{-\frac{1}{2\mu} (B^{-1}k + k^{-1} )}
\end{eqnarray}
Here the last line is obtained via Taylor series.  In addition, the factor $e^{-\frac{1}{2\mu k}}$ approaches $1$ as $k \rightarrow \infty$ and has a minor effect on the integral.  We therefore approximate the wave as
\begin{eqnarray} \label{LargeMu_h}
	h(\mathbf{x},  t_{n+1}; \mathbf{y}_n) &\approx& -\int \frac{1}{4\pi\mu k} e^{-\frac{1}{2\mu B}k}e^{\imath \mathbf{k}(\mathbf{x} - \mathbf{y}_n)} \du \mathbf{k} + O(\mu^{-3/2}) \\
	&=& -\frac{1}{4\pi\mu} \int_{0}^{2\pi} \int_{0}^{\infty} e^{ (-\frac{1}{2\mu B} +\imath r \cos \theta) k} \du k \du \theta + O(\mu^{-3/2}) \\
 	&=& -\frac{1}{2\pi} \int_{0}^{2\pi}  \frac{\du \theta}{(B^{-1} - \imath 2\mu  r \cos \theta)} + O(\mu^{-3/2}).
\end{eqnarray}
Here, we have aligned the droplet position and velocity with the x-axis ($\mathbf{y}_n = r_n \hat{\mathbf{x}}$, $\mathbf{v}_n = v_n \hat{\mathbf{x}}$) and introduced $r = |\mathbf{x} - \mathbf{y}_n|$.  To compute the last integral, we make the change of variables $z = e^{\imath \theta}$, and proceed by evaluating the residues enclosed by the unit circle $|z| = 1$.  For large $\mu$ we have the over damped wave:
\begin{eqnarray} \label{LargeMu_h2}
	h(r, t_{n+1}) &=& -\frac{1}{\sqrt{B^{-2} + 4 (\mu r)^2}} + O(\mu^{-3/2}).
\end{eqnarray}
Taking $\partial_r h(r)$, we have
\begin{eqnarray} \label{LargeMu_Map}
	v_{n+1} = (1-\gamma) v_n - \frac{4\mu^2 F v_n}{(B^{-2} + 4\mu^2 v_n^2)^{3/2}}.
\end{eqnarray}
The iterative map (\ref{LargeMu_Map}) has only one fixed point $v_{n} = 0$, regardless of $F$ and $B$. Hence, the strongly over-damped waves do not produce a bifurcation to walking motion as one increases $F$.  Physically, the effect of damping smooths out the wave curvature, thereby inhibiting a transition to stable walking.

\begin{figure}[htb!]
	\centering
    \includegraphics[width = 0.6\textwidth]{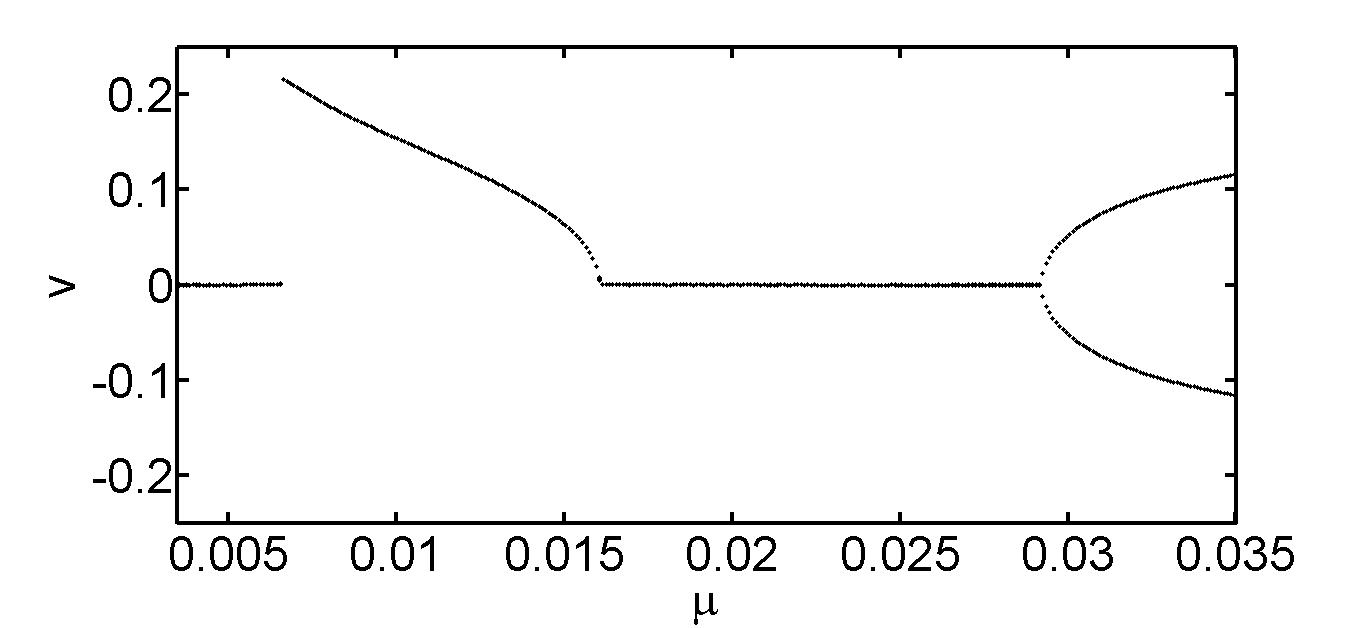} \\
    \caption{Plot shows the droplet velocity for varying $\mu$ at a fixed $F/F_{crit} = 1.1$.  The parameters are $B = 120$ and $\gamma = 1$ while $F_{crit}$ is the critical forcing for $\mu = 0.008$. At small $\mu < .006$ the fixed forcing $F$ is not large enough to induce walking.  At large $\mu$ the droplet does not walk but rather oscillates about a fixed point.}
 \label{VaryMu}
\end{figure}

To illustrate the effect of $\mu$ over typical experimental values ($.001$ to $.016$), figure (\ref{VaryMu}) shows the droplet velocity for a fixed $F$ and different values of $\mu$.  Over most experimental values, the velocity has a non-zero fixed point, or walking solution.  As $\mu$ increases, the fixed point $v = 0$ first becomes a stable attractor (around $\mu \sim 0.016$) indicating a reverse transition from walking to bouncing. At larger $\mu$, $v = 0$ destabilizes into a periodic orbit at which point the droplet bounces back and forth about one fixed location in space. Over experimental values, $\mu$ does not change the qualitative walking behavior, and only has a minor effect on the velocity of the droplet.

\section{General requirements for stable walking}

In this section we examine a set of general requirements for the map (\ref{Update_y})--(\ref{Update_v}) and assumption (A3) to yield stable walking solutions in free space ($\Omega = \mathbbm{R}^2$).

Firstly, a droplet impact at position $\mathbf{y}_n$ will generate a radially symmetric wave field about $\mathbf{y}_n$.  Secondly, we note that the translational symmetry in free space implies that a wave field generated by an impact at $\mathbf{y}_n$ will only depend on the difference $(\mathbf{x} - \mathbf{y}_n)$.  Hence, $h(\mathbf{x}, t; \mathbf{y}_n)$ has the general form
\begin{eqnarray}
    h(\mathbf{x}, t; \mathbf{y}_n) = h( |\mathbf{x} - \mathbf{y}_n|, t).
\end{eqnarray}
Letting $r = |\mathbf{x} - \mathbf{y}_n|$, we can introduce the function $g(r) = -\partial_r h(r, 1)$ where $h(r, 1)$ is the wave field generated by an impact at time $t = 0$ and evaluated at one strobe period later (ie. $t = 1$).  Physically, $g(r)$ describes the radial forcing on the droplet in the iterative map (\ref{Update_y})--(\ref{Update_Map}).  Using the fact that $\mathbf{y}_{n+1} - \mathbf{y}_n = \mathbf{v}_n$, we then obtain a one dimensional, iterative map for the droplet velocity in the radial direction
\begin{eqnarray} \label{General_Map}
    v_{n+1} = (1-\gamma) v_n + F g(v_n).
\end{eqnarray}
Without loss of generality, we align the system with the x-axis (ie. $\mathbf{v}_n = v_n \mathbf{\hat{x}}$) and further assume the parameters $0 < \gamma \leq 1$ and $F \geq 0$.  Since $h(r, t)$ is radially symmetric, it follows that $h(r, t)$ is an even function of $r$.  Therefore, (provided $h(r, t)$ is regular at $r = 0$), $g(v)$ is an odd function of $v$ and $g(0) = 0$.  Hence, (\ref{General_Map}) always admits $v = 0$ as a fixed point.

For stable walking solutions, we require the existence of a nonzero stable fixed point.  The following criteria guarantee such a point.  Let $v^* > 0$ and satisfy the following propositions
\begin{itemize}
\item[P1.] Existence of a nonzero fixed point
\begin{eqnarray}
    g(v^*) > 0. \label{ExistenceFP}
\end{eqnarray}
\item[P2.] Stability of the fixed point
\begin{eqnarray} \label{StabilityFP}
    0 < 1 - \frac{v^* g'(v^*)}{g(v^*)} < \frac{2}{\gamma}.
\end{eqnarray}
\end{itemize}
Here item (\ref{ExistenceFP}) implies taking a forcing $F = \gamma v^*/ g(v^*) > 0$ yields the fixed point velocity $v^*$.  Physically, condition (\ref{ExistenceFP}) guarantees that the wave propels the droplet forward at each interaction.  Meanwhile, condition (\ref{StabilityFP}) guarantees the stability of the linearized map at $v = v^*$.  Practically, one may simply plot the function $f(v) = 1 - \frac{v g'(v)}{g(v)}$ to determine whether the corresponding wave field supports a stable walking droplet.  Lastly, conditions (\ref{ExistenceFP})--(\ref{StabilityFP}) only guarantee a stable walking solution.

An additional, yet independent criteria for a bifurcation from stable bouncing to walking motion is an instability at $v = 0$
\begin{itemize}
\item[P3.] Instability of the $v = 0$ fixed point
\begin{eqnarray} \label{InstabilityFP}
   g'(0) > 0
\end{eqnarray}
\end{itemize}
Here the condition (\ref{InstabilityFP}) guarantees that taking $F > \gamma/g'(0)$ yields an unstable fixed point at $v = 0$\cite{FixedPoint}.  Physically, the condition (\ref{InstabilityFP}) corresponds to a concave down wave field $h''(0) < 0$ and assures that the instability will propagate the droplet in one direction.

One should note that the criteria (\ref{ExistenceFP})--(\ref{StabilityFP}) yields a stable walking solution provided $F = \gamma v^*/ g(v^*) > 0$, while (\ref{InstabilityFP}) is a separate condition which guarantees that the fixed point $v = 0$ becomes unstable for $F > \gamma/g'(0)$.  In general, the simultaneous stability or instability of the $v = 0$ and $v = v^*$ fixed points depends on other details of the wave field.  For example, one may have both subcritical and supercritical pitchfork bifurcations depending on the sign $h^{(4)}(0)$.
We also remark that another standard wave field that fails condition (\ref{ExistenceFP}) is the linear Green's function solution to the Helmholtz equation.  Such a model is proposed in \cite{CouderFort2006}, however, they obtain walking solutions through the summation of many past bounces.

\section{Movement in a square} \label{BoundedDomainA}

In this section we examine the movement of droplets in a square domain using the model (\ref{Update_y})--(\ref{Update_v}).  To capture the reflection of the fluid waves against the wall, we impose a Neumann boundary condition on the velocity potential $\frac{d \phi}{d\mathbf{n}} = 0$ corresponding to a no fluid flux boundary condition on the bath.  Equivalently, such a condition corresponds to a Neumann boundary condition on $h(\mathbf{x}, t; \mathbf{y}_n)$:
\begin{eqnarray} \label{PhiBC}
    \frac{dh}{d\mathbf{n}} = 0, \hspace{4mm} \mathbf{x} \in \partial \Omega
\end{eqnarray}
where $\mathbf{n}$ is the unit normal along the domain boundary.  For instance, differentiating equation (\ref{DimensionlessEq1}) and projecting onto the boundary yields $\partial_t \frac{dh}{d\mathbf{n}} = \partial_z\frac{d\phi}{d\mathbf{n}} = 0$.  Hence the boundary condition $\frac{d \phi}{d\mathbf{n}} = 0$ implies $\frac{dh}{dz} = C$, a constant in time.  Since the constant $C = 0$ at $t = 0$, we take $\frac{dh}{d\mathbf{n}} = 0$ for all time.

\begin{figure}[htb!]
	\centering
    \includegraphics[width = 0.6\textwidth]{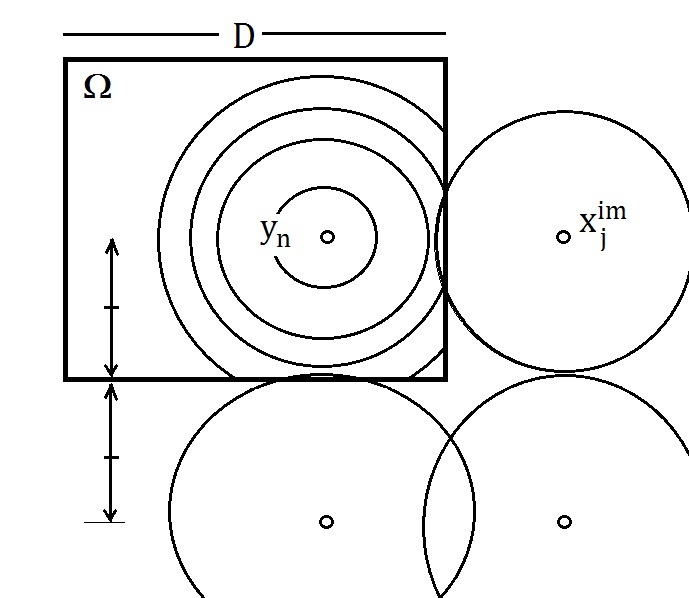} \\
    \caption{Plot shows the closest image points used to compute the wave field in a square domain.  Each image point acts as a source with wave field $h_{FS}(r)$.}
 \label{MethodImages}
\end{figure}

To evaluate the motion of the droplet in a square domain, we must compute the wave field $h(\mathbf{x}, t)$ at each iteration of the map.  To aid in the computation of the field, we may exploit the method of images \cite{TikhonovSamarskii1963} and the geometry of a square.  For instance, since the field $h(\mathbf{x}, t)$ satisfies Neumann boundary conditions, and is generated by an impulse at each step, the solution may be generated by an infinite array of image points of the free space wave field $h_{0}(r, t)$ at properly chosen locations $\mathbf{x}_j^{im}$.  The location of the image points depend on $\mathbf{y}_n$, the closest ones being at points reflected across the domain wall boundaries.  One may then compute the wave field from the knowledge of the free space wave
\begin{eqnarray} \label{WaveSuperposition}
	h(\mathbf{x}) &=& \sum_{m = 0}^{n-1}\Big( h_0(|\mathbf{x}-\mathbf{y}_{n-m}|, m+1) + \sum_j h_{0}(|\mathbf{x} - \mathbf{x}_{j}^{im}(\mathbf{y}_m)|, m+1) \Big).
\end{eqnarray}
Here the  $\mathbf{x}_{j}^{im}(\mathbf{y}_m)$ are the image points $\mathbf{x}_j^{im}$ which depend on the source term $\mathbf{y}_m$.  Figure (\ref{MethodImages}) illustrates the wave field with the most important image points.  The addition of the image points yield correct boundary conditions for $h(\mathbf{x})$ on the bottom and right side of the square.

To simplify the expression (\ref{WaveSuperposition}), we may separate out the free space bounces at locations $\mathbf{y}_{n-j}$ as
\begin{eqnarray}
	h_{FS}(\mathbf{x}) &=& \sum_{m = 0}^{n-1} h_0(|\mathbf{x}-\mathbf{y}_{n-m}|, m+1)
\end{eqnarray}
For simplicity, we may then approximate $h_{FS}(\mathbf{x})$ as a radially symmetric ansatz centered around the most recent position
\begin{eqnarray} \label{WaveAnsatz}
	h_{FS}(r) &\approx& 20 J_0(11.5 r) e^{-1.15 r} \\
	r &=& |\mathbf{x} - \mathbf{y}_n|.
\end{eqnarray}
Figure (\ref{FullWaveField}) compares the approximation (\ref{WaveAnsatz}) to the fully developed gravity-capillary wave in the radial direction.  The approximation here is also similar to the phenomenological ansatz provided in \cite{CouderFort2006, FortEddiBoudaoudMoukhtarCouder2010}.

In making the approximation (\ref{WaveAnsatz}), one is effectively concentrating all previous bounces onto $\mathbf{y}_n$.  As a result, the approximation captures contributions from previous impacts, however suppresses all memory effects.  Again, such an approximation is valid at low forcing far from the Faraday threshold.  The ansatz also neglects the Doppler effect present from a moving source. Experimentally, however, the Doppler effect is negligible at low forcing since the droplet velocity is small compared to the group velocity of the wave (ie. the ratio is $\sim 0.06$).  In concentrating all impacts onto the previous location, (\ref{WaveAnsatz}) greatly simplifies the iterative map, and may aid in future work on developing evolution equations for the droplet probability distribution.

Finally, since $h_{FS}(r)$ decays quickly, one may truncate the sum (\ref{WaveSuperposition}) for the efficient computation of the wave field.  In our case we keep the first order contributions as illustrated in figure (\ref{MethodImages}).  The field $h(\mathbf{x})$ then becomes
\begin{eqnarray} \label{WaveSuperposition2}
	h(\mathbf{x}) &=& h_{FS}(|\mathbf{x}-\mathbf{y}_{n}|) + \sum_j h_{FS}(|\mathbf{x} - \mathbf{x}_{j}^{im}(\mathbf{y}_n)|) \Big)
\end{eqnarray}

\subsection{Large Domain} \label{LargeDomain}

In this section we examine solutions to the map (\ref{Map_y})--(\ref{Map_v}) where the wave field is given by (\ref{WaveAnsatz})--(\ref{WaveSuperposition2}).
In our numerical evaluation of the map, we fix $\gamma = 1$ and the size of the box $D = 12$ (approximately $20 cm$) to be much larger than one wavelength.  We then examine trajectories for different forcing.  Since we model the droplet interaction with the boundary of the domain entirely by reflected waves, at large forcing there is a possibility that the droplet may physically collide or jump over the boundary.  We therefore limit our attention to parameters which yield bounded trajectories ($F/F_{crit} < 0.869$ where $F_{crit} = 1150^{-1}$ is a normalized forcing from the gravity-capillary model), namely those which reflect off the walls.

\begin{figure}
  \centering
  \subfloat[$\gamma = 1$, $F/F_{0} = .65$]{\label{CircularOrbit}\includegraphics[width=0.4\textwidth]{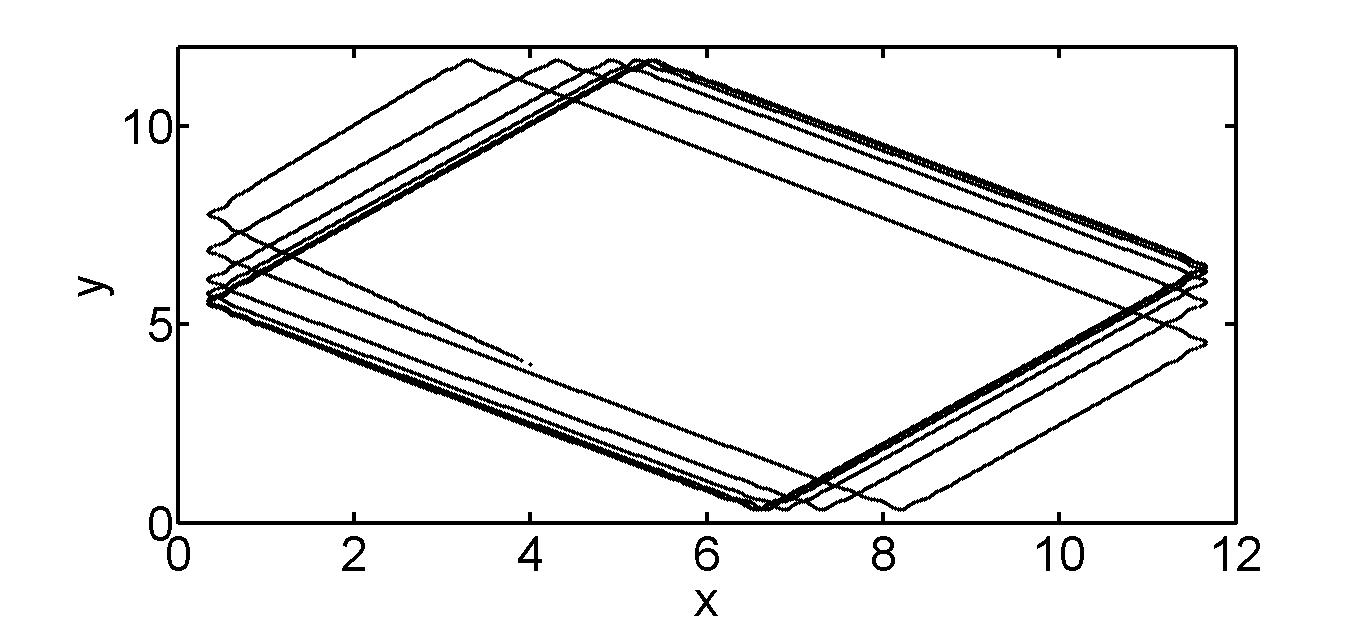}}
  \subfloat[$\gamma = 1$, $F/F_{0} = .75$.]{\label{DenseTrajectory}\includegraphics[width=0.4\textwidth]{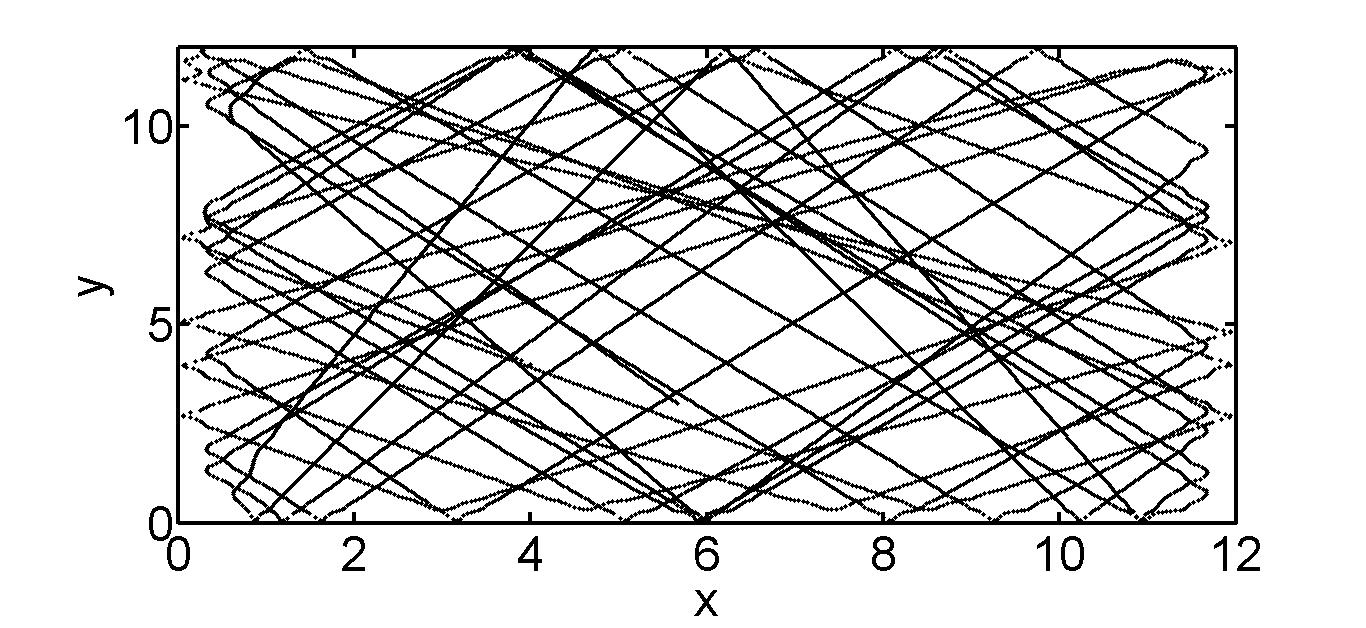}}
  \caption{The long time spatial trajectories for $D = 12$ ($\sim 20 cm$) collapse into (a) quasiperiodic orbit at lower forcing, or (b) travel throughout the domain at large forcing. Here $F_{crit} = 1150^{-1}$ is a normalization force from the free space walking threshold.}
\end{figure}

Depending on the parameters of the underlying wave field, the long time behavior of particle trajectories may be classified into two categories depending on the nature of the limiting set: those which approach a circular quasiperiodic orbit ($0.610 < F/F_{crit} < 0.733$) , and those which continually traverse the domain ($0.733 < F/F_{crit} < 0.869$).  In the second category, the trajectories appear to form a dense set throughout the spatial domain.  To illustrate the different scenarios, figure (\ref{CircularOrbit}) shows the path of a droplet approaching a quasiperiodic orbit while figure (\ref{DenseTrajectory}) shows part of a dense trajectory.  The emergent pattern, however, is not related to cavity modes of the square, but rather results because droplet trajectories tend to travel along paths near angles of $\pi/4$ with respect to the x-axis.  For instance, although not shown, the statistics of the droplet velocity angles are centered around angles of $\pm \pi/4$.  For box sizes much larger than the natural wavelength of $h(\mathbf{x})$, the droplet behaves vary much like an isolated particle.  When the droplet approaches a wall, the droplet reflects off the wall through the mediation of the reflected wave field.  The reflection is somewhat analogous to a billiard ball on a table since the incident and reflected angles are approximately equal.

\subsection{Small domain}

In this section, we examine the long time behavior of droplet trajectories for a domain size comparable to the fluid wavelength (ie. $D \sim 1-2 cm$).  Although such domains are experimentally small compared to current setups, they correspond to the classical analogy of having a quantum system with the de Broglie wavelength comparable to the domain size. Unlike the previous section, in small domains the wave field has time to respond to the geometry of the box.
Again we work well below the Faraday threshold and neglect strong memory effects where such interactions can lead to additional droplet dynamics, even in large cavities $D \gg \lambda_f$.

To determine the long time behavior of the map, we fix a set of parameter values and examine the trajectories for many different initial conditions.  The data for each initial condition is chosen to survey the phase space within a bounded set by prescribing, $|\mathbf{v}_0| < 0.25$, and taking a maximum distance between $\mathbf{y}_0$ and the nearest wall to be less than $0.1$.  Here the exact bounds of $0.25$ and $0.1$ are chosen somewhat arbitrarily to include a large, physically relevant, region of phase space.

\begin{figure} [htb!]
  \centering
  \subfloat[F = 0.22]{\includegraphics[width=0.4\textwidth]{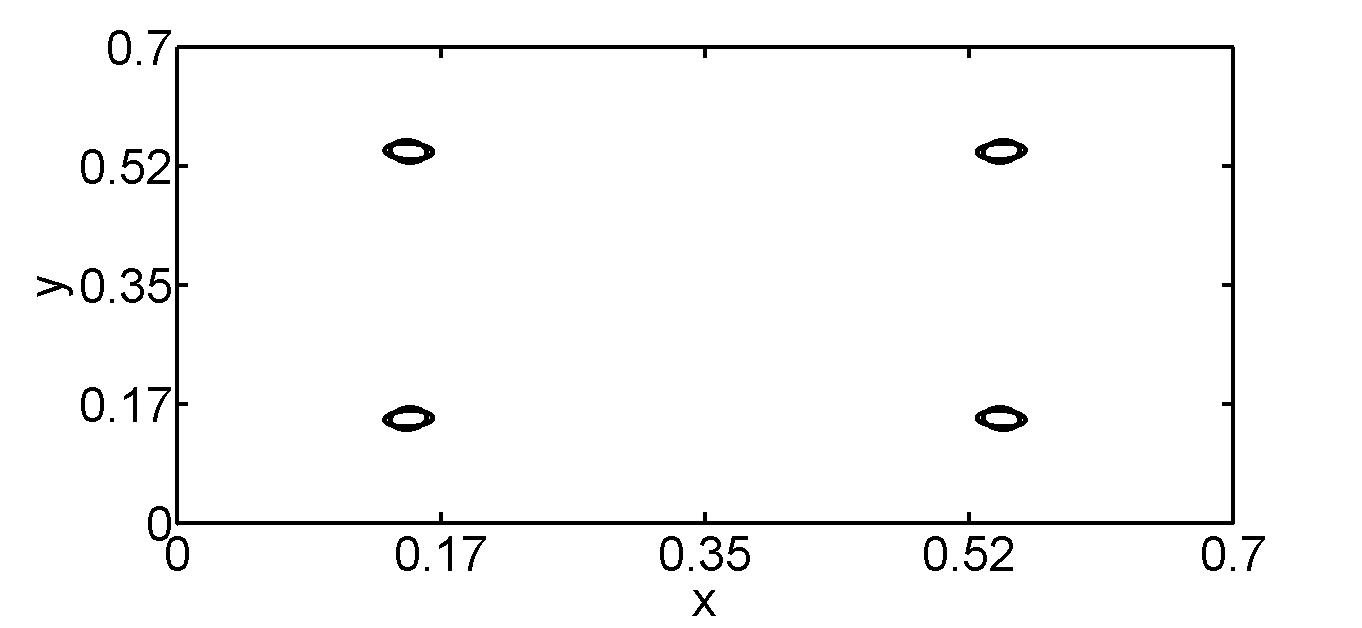}}
  \subfloat[F = 0.43]{\includegraphics[width=0.4\textwidth]{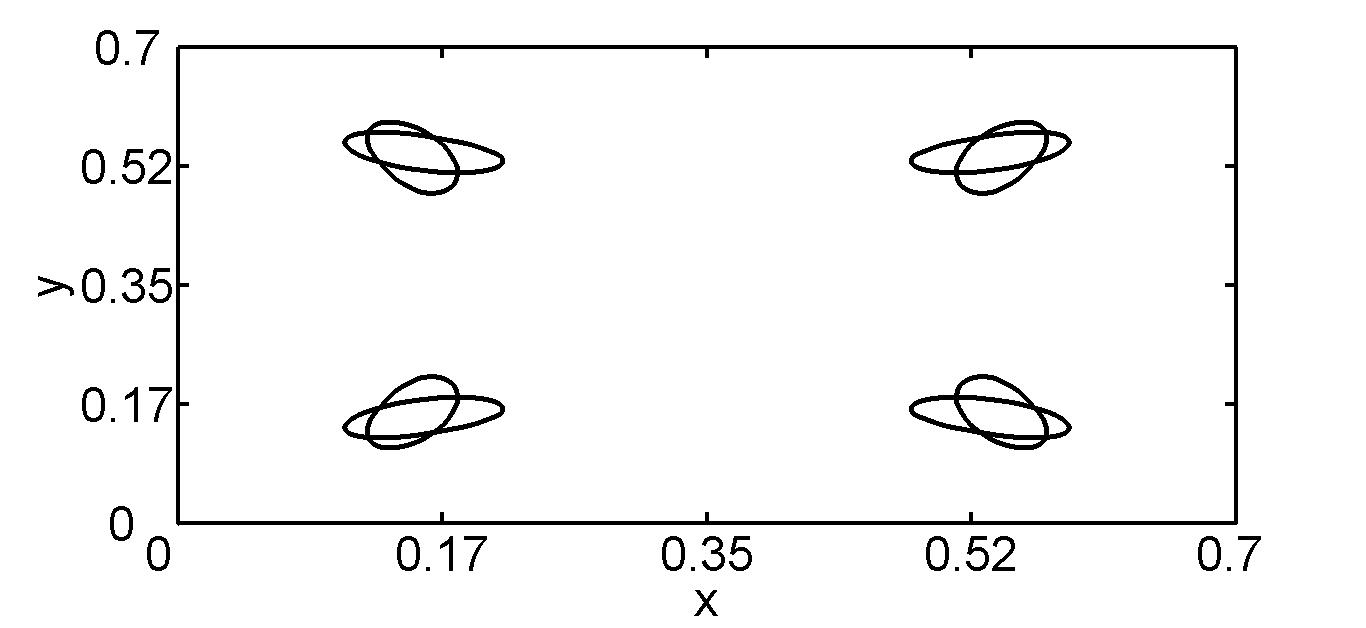}} \\
\subfloat[F = 0.48]{\includegraphics[width=0.4\textwidth]{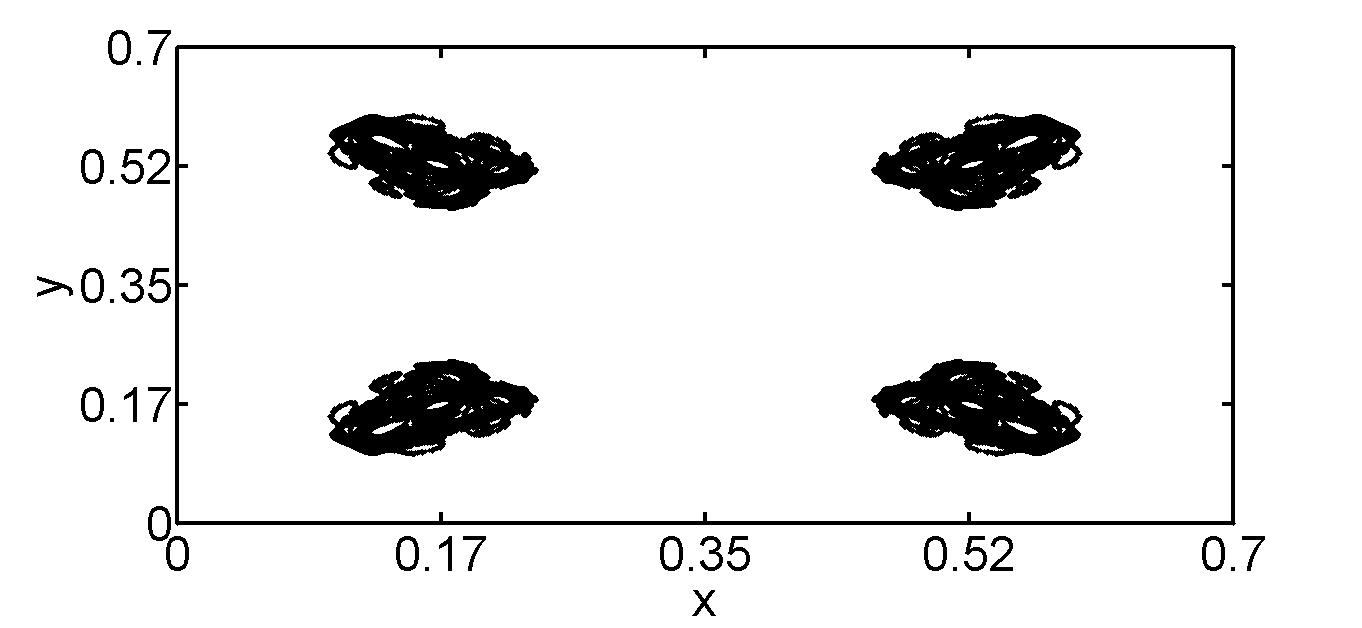}}
  \subfloat[F = 0.52]{\includegraphics[width=0.4\textwidth]{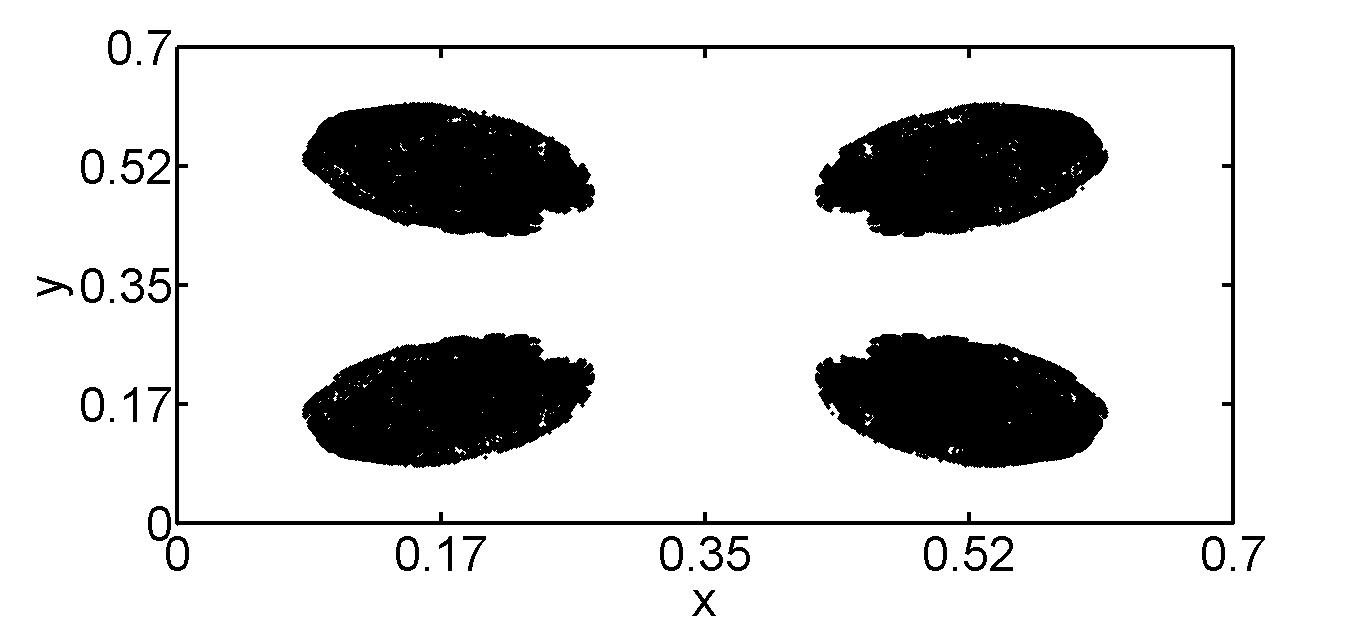}}\\
  \subfloat[F = 0.57]{\includegraphics[width=0.4\textwidth]{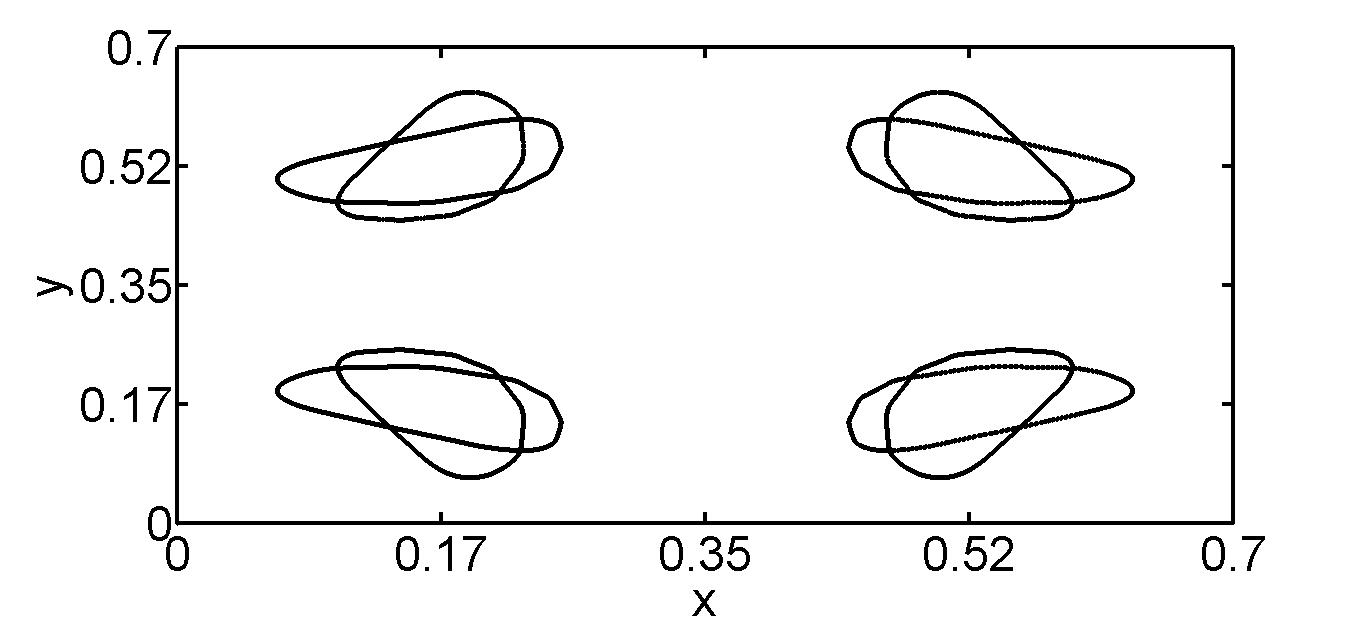}}
\subfloat[F = 0.61]{\includegraphics[width=0.4\textwidth]{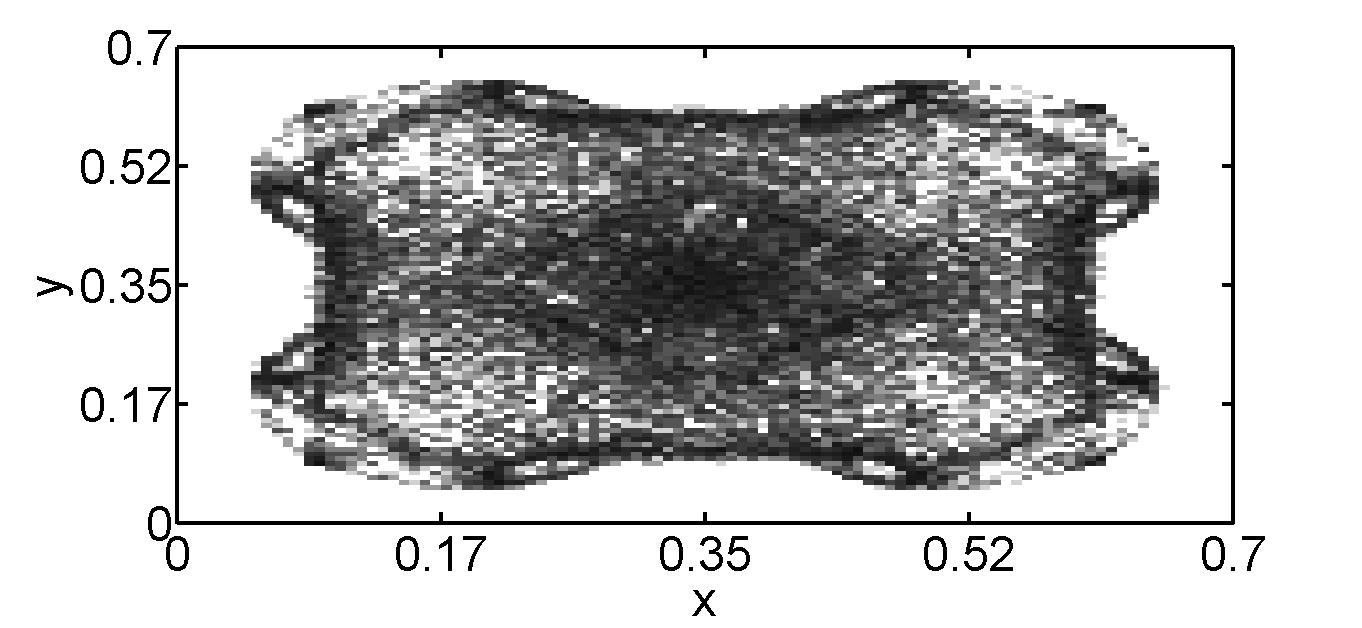}}
  \caption{Attracting sets for the model response (\ref{WaveAnsatz}) and various forcing where $D = 0.7$.}
  \label{fig:Box8}
\end{figure}

For each initial condition, we remove any transient effects by first evolving the trajectory for several thousand iterations.  After discarding the transients, we then evolve the droplet for several thousand more iterations, and project the trajectory from the four dimensional phase space $(\mathbf{x}, \mathbf{v})$ onto the two dimensional physical domain $(\mathbf{x})$.  In many cases the trajectory approaches an attracting set in the form of a periodic or quasiperiodic orbit.  We now describe in detail the long time trajectories as one varies $F$ for fixed $D$.  Specifically, we consider in detail the case of $D = 0.7$ ($\sim 1.1 cm$), which corresponds to a box size of roughly one wavelength, and $D = 1.05$ ($\sim 1.7 cm$) which is just under two wavelengths of the wave field.

Initially, with small values of the forcing $F$, the long time trajectories approach one of several quasiperiodic orbits.  These orbits form attracting sets for different regions of phase space. For instance, the exact orbit a trajectory approaches depends only on the trajectories initial conditions. Together the collection of all attracting orbits form a symmetric array on which the precise pattern depends on the parameters $F$ and the box size $D$.  Qualitatively the number of orbits depends most strongly on the box size $D$.  The reason is that the droplets tend to localize near the troughs from the waves reflected off the domain boundaries.  The larger box sizes allow for more wavelengths from the reflected waves. For instance, over a wide range of forcing $F$, a box size of $D = 0.7$ supports 4 quasiperiodic attracting regions, while $D = 1.05$ contains 8.  Here the shape of the array, and number of orbits appear linked to the geometry of the domain.

\begin{figure}[htb!]
  \centering
  \subfloat[F = 0.11]{\includegraphics[width=0.4\textwidth]{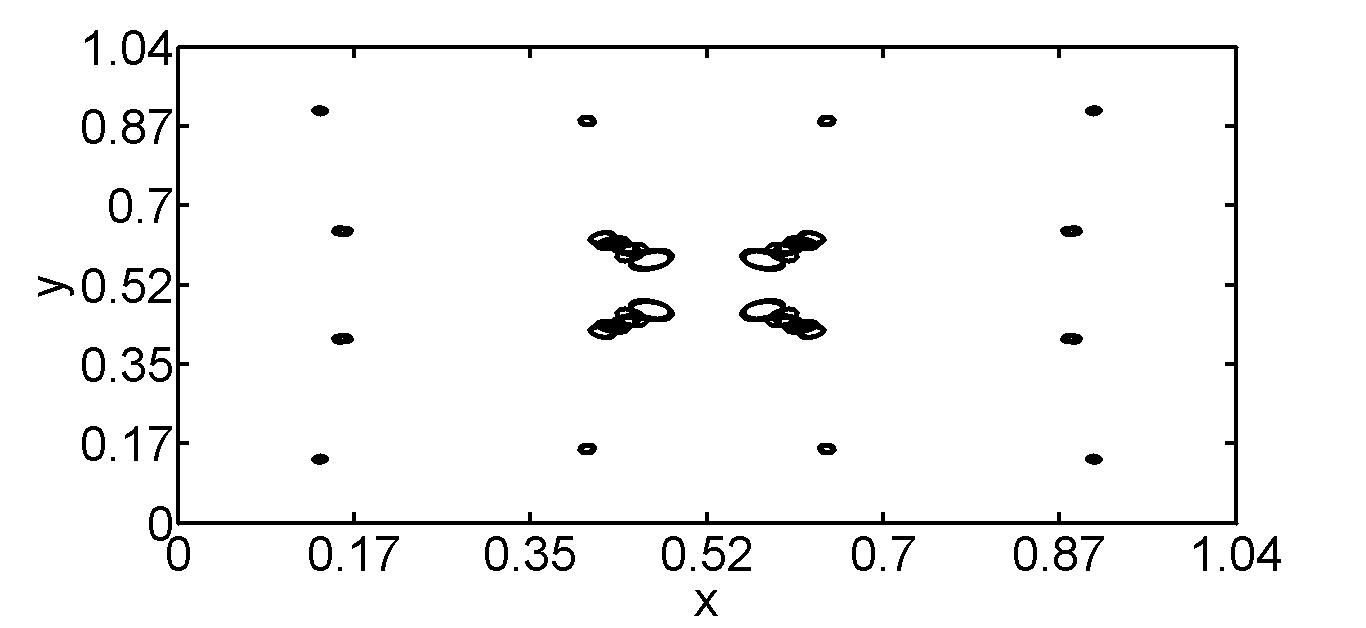}}
  \subfloat[F = 0.22]{\includegraphics[width=0.4\textwidth]{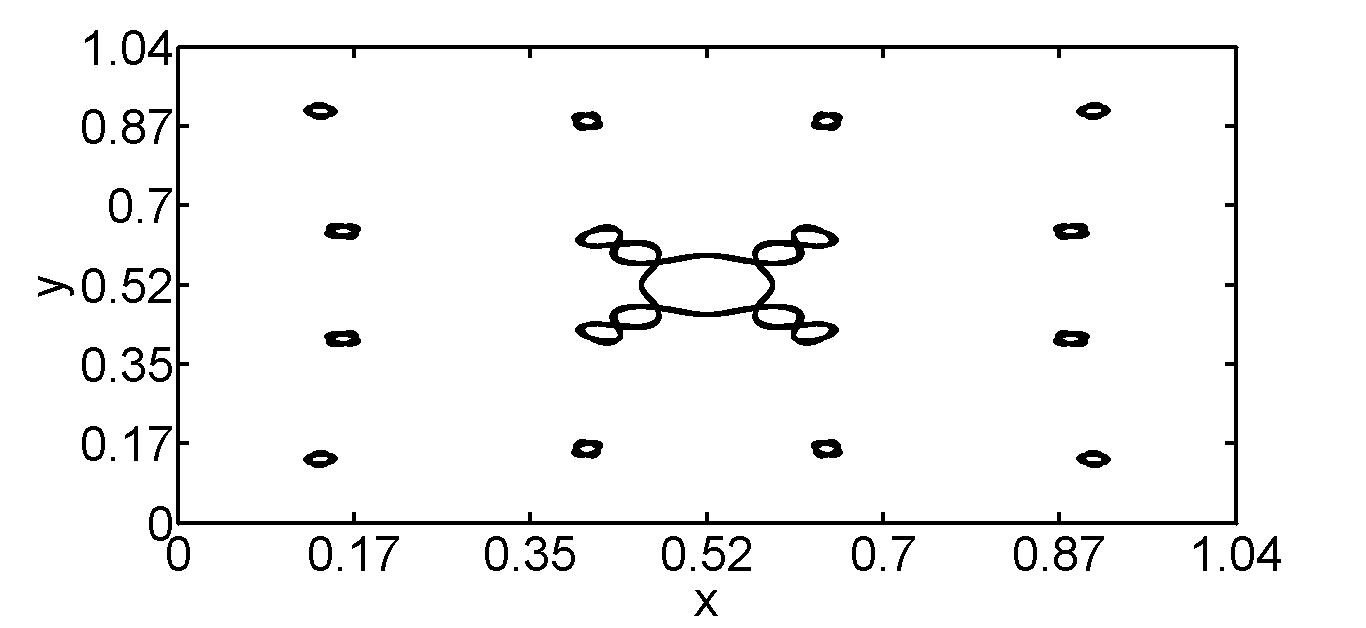}} \\
  \subfloat[F = 0.43]{\includegraphics[width=0.4\textwidth]{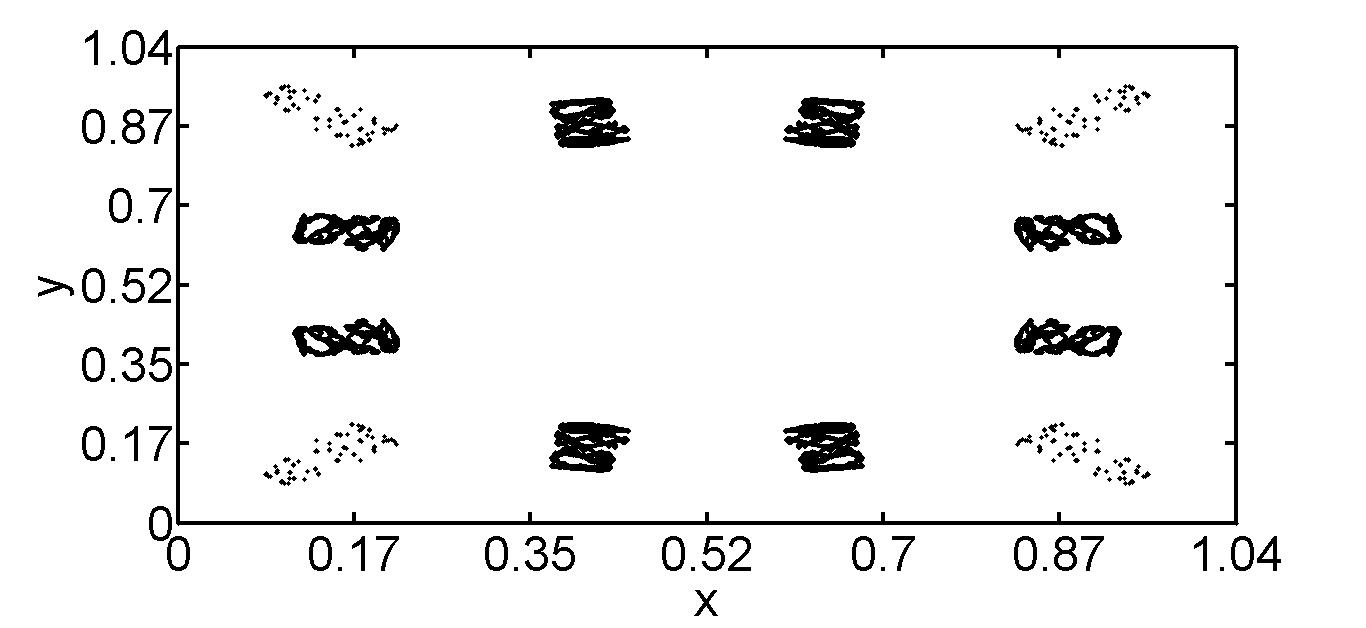}}
  \subfloat[F = 0.54]{\includegraphics[width=0.4\textwidth]{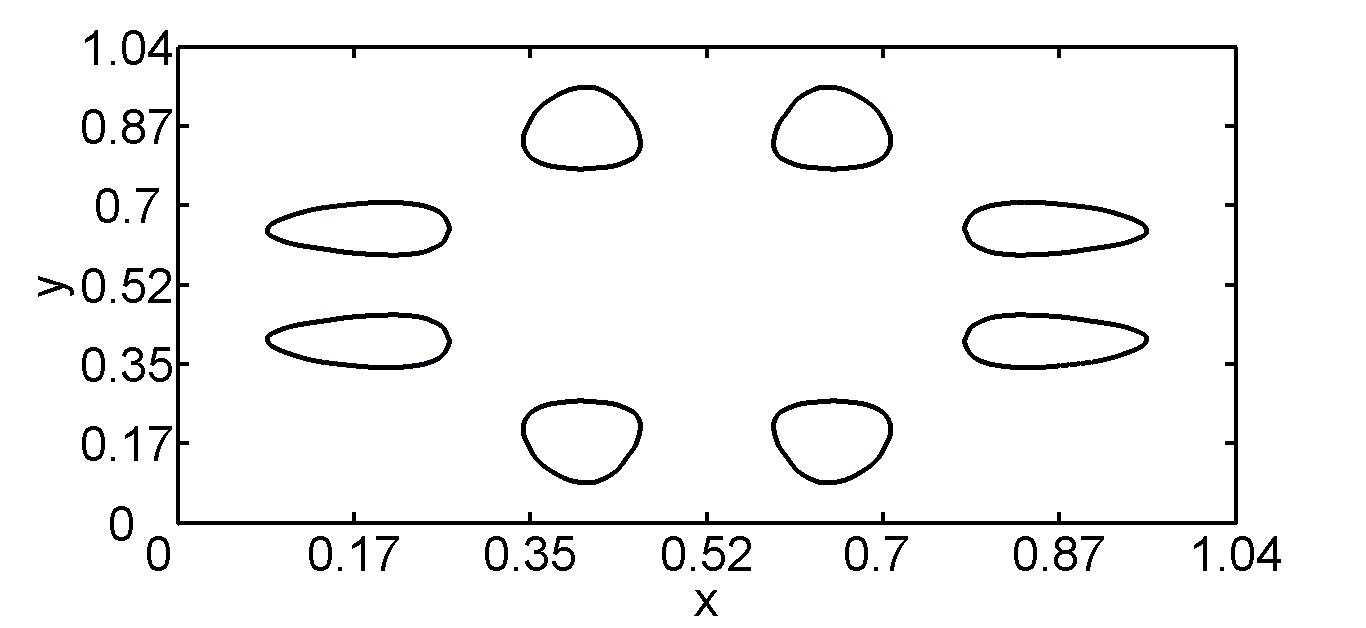}} \\
\subfloat[F = 0.65]{\includegraphics[width=0.4\textwidth]{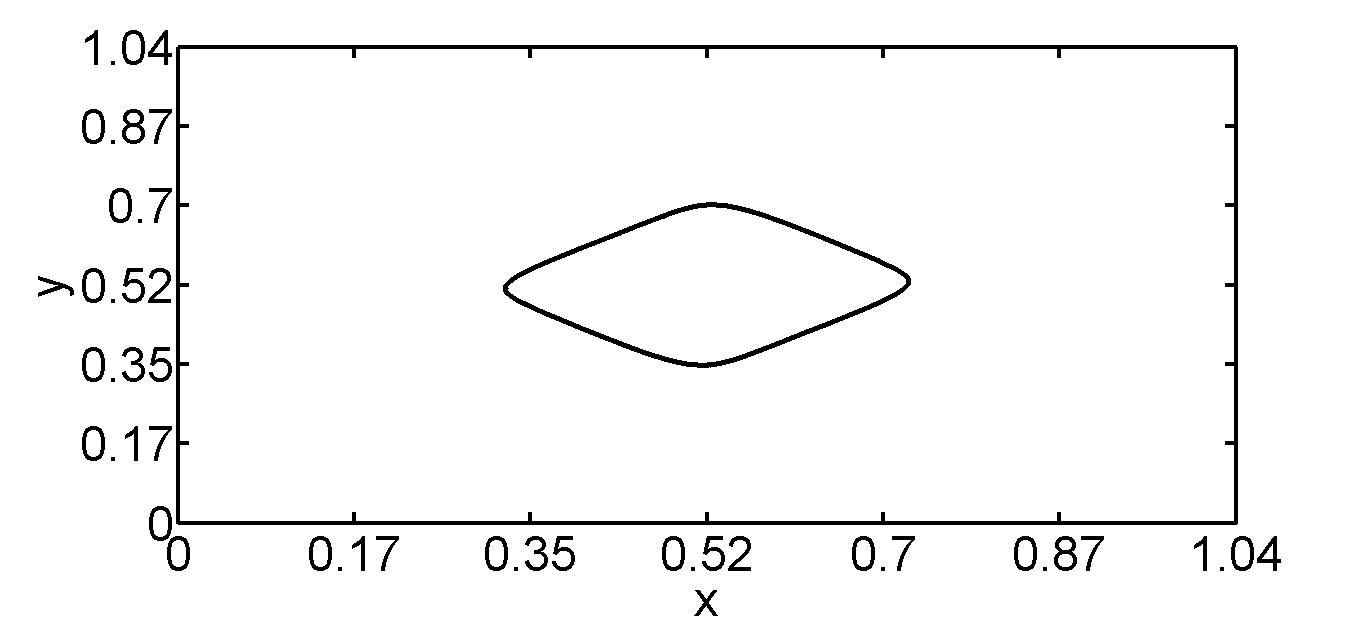}}
  \subfloat[F = 0.66]{\includegraphics[width=0.4\textwidth]{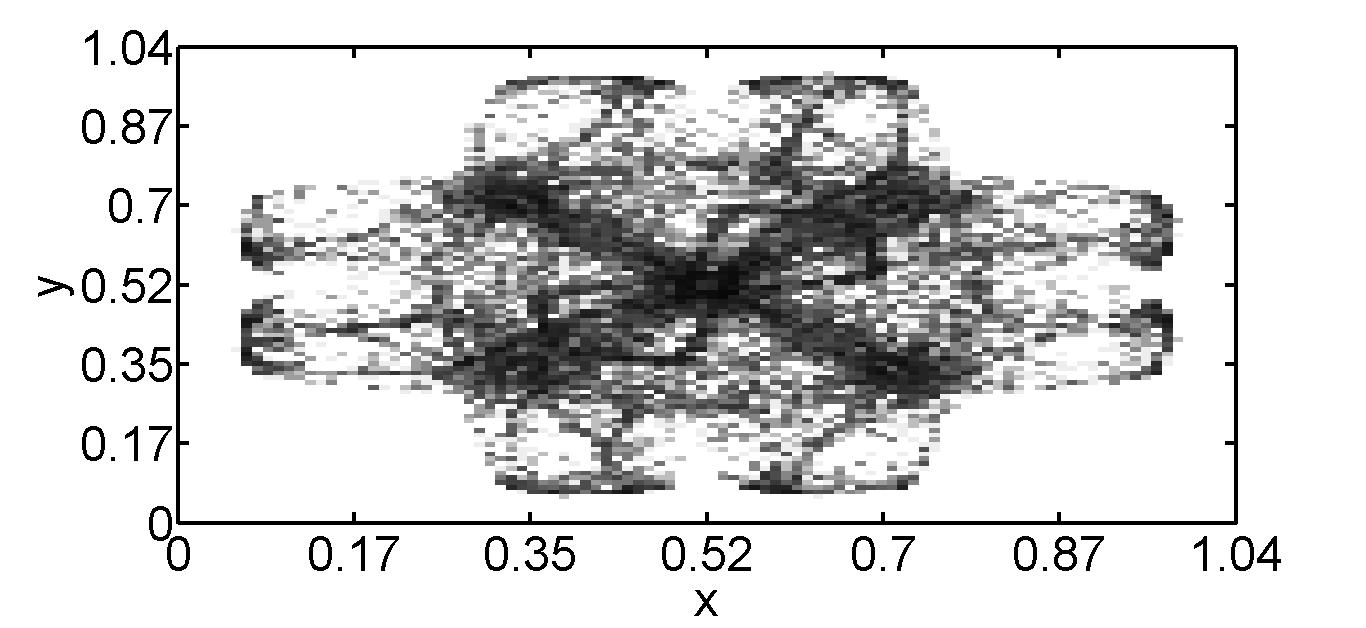}}
  \caption{Attracting sets for the model response (\ref{WaveAnsatz}) and various forcing where $D = 1.05$.}
  \label{fig:Box12}
\end{figure}

As the forcing increasing, the spatial radii of the quasiperiodic orbits grow.  For instance figures (\ref{fig:Box8}) and (\ref{fig:Box12}) show the attracting sets for box sizes $0.7$ and $1.05$ with different forcing.  At sufficient forcing, the nature of the attracting sets change from thin circular orbits to thick, sets.  At large forcing, the localized attracting sets break down, and the droplet wonders throughout the domain.  Figures (\ref{fig:Box8}f) and (\ref{fig:Box12}f) show a shaded probability distribution for the droplets position at the large forcing.  Despite the fact that there are no longer circular quasiperiodic orbits, the droplet still spends a significant time near the former quasiperiodic orbit regions.  For instance, there are similarities in the dark outlines of (\ref{fig:Box8}e) and (\ref{fig:Box8}f), as well as (\ref{fig:Box12}d) and (\ref{fig:Box12}f).  Although the distribution shows dark, highly traversed regions, and light, vacated regions in a regular array that appears related to the underlying wave field, the exact dependence is not completely understood.  Here we defer further investigation to future work.  For instance, in future work we seek to examine the relation between the dynamical systems invariant measure, and the underlying wave field.

\subsection{Conclusions}
Through the introduction of an iterative map, we model the dynamics and trajectories of bouncing droplets on an oscillating fluid bed.  As a first step, we examine the droplets bifurcation from a stable bouncing state to a stable walking one.  In addition, we list several requirements for the underlying wave field to undergo such a bifurcation.  Using the map, we then investigate the droplet trajectories for wave responses in a square (billiard ball) domain.
In the case of a large domain, we recover limit cycle and dense trajectories which appear similar to those reported in \cite{CouderFort2006}.
Lastly, in small domains we show that for low forcing, trajectories tend to approach circular attracting sets.  As one increases the forcing, the attracting sets break down and the droplet tends to travel through space, jumping between the former attracting regions.
In future work we plan to further examine the statistical nature of the droplet trajectories, including their transport properties and invariant measures.

\subsection{Acknowledgments}

The author would like to thank John Bush for originally posing the problem of understanding the bouncing droplet trajectories.  The author has also vastly benefited from many conversations with Renato Calleja, Tristan Gilet, Anand Oza, Jean-Christophe Nave and Ruben Rosales.  Lastly, the author gratefully acknowledges the many helpful comments of an anonymous reviewer. The work was partially supported by NSERC and NSF grant $DMS�0813648$.

\newcommand{\noopsort}[1]{} \newcommand{\printfirst}[2]{#1}
  \newcommand{\singleletter}[1]{#1} \newcommand{\switchargs}[2]{#2#1}


\begin{thebibliography}{10}
\bibitem{LengthScale}
We remark that another natural (Faraday) length scale is set using the linear dispersion relation $\omega^2 = g k + \sigma k^3/\rho$ in combination with the period $T$.  Such a choice, however leads to a more complicated form for the coefficients in the dimensionless equation.
\bibitem{FixedPoint}
We omit the case of an oscillating instability when the derivative of the iterative map is less than $-1$.
\end{thebibliography}
\end{document}